\documentclass[letterpaper,twocolumn,twoside]{IEEEtran}

\usepackage{multirow}
\usepackage{amsmath,graphicx}
\usepackage{latexsym,amssymb}
\usepackage{url}
\usepackage{relsize}
\usepackage{bbm}
\usepackage{cite}

\newcommand{\argmax}{\operatornamewithlimits{arg\,max}}
\newcommand{\argmin}{\operatornamewithlimits{arg\,min}}

\newlength{\mylength}
\newlength{\newlengthtwo}
\newlength{\barlength}
\newcommand {\balpha}{\ensuremath \boldsymbol{\alpha}}
\newcommand {\bz}{\ensuremath {\mathbf z}}

\title{Photon-Efficient Computational 3D and \\
   Reflectivity Imaging with Single-Photon Detectors
\thanks{This material is based upon work supported in part by the
   US National Science Foundation under Grant No.~1161413,
   a Samsung Scholarship, and a Microsoft PhD Fellowship.}
\thanks{This work will be presented in part at the IEEE International
   Conference on Image Processing, Paris, France, October 2014.}
}

\author{Dongeek Shin,
        Ahmed Kirmani,
        Vivek K Goyal, and
        Jeffrey H. Shapiro
\thanks{D. Shin, A. Kirmani, and J. H. Shapiro are with the 
  Department of Electrical Engineering and Computer Science,
  Massachusetts Institute of Technology, Cambridge, MA 02139 USA\@.}
\thanks{V. K. Goyal is with the
  Department of Electrical and Computer Engineering, Boston University,
  Boston, MA 02215 USA\@.}
}

\begin{document}

\maketitle

\begin{abstract}
Capturing depth and reflectivity images at low light levels from
active illumination of a scene has wide-ranging applications.
Conventionally, even with single-photon detectors, hundreds
of photon detections are needed at each pixel to mitigate
Poisson noise. We develop a robust method for estimating
depth and reflectivity using on the order of 1 detected photon per pixel
averaged over the scene.  Our computational imager combines physically
accurate single-photon counting statistics with
exploitation of the spatial correlations present in real-world reflectivity
and 3D structure. Experiments conducted in the presence of strong
background light demonstrate that our computational imager is able
to accurately recover scene depth and reflectivity, while traditional
maximum-likelihood based imaging methods lead to estimates that
are highly noisy.
Our framework increases photon efficiency 100-fold over traditional processing
and also improves, somewhat, upon first-photon imaging under a
total acquisition time constraint in raster-scanned operation.
Thus our new imager will be useful for rapid, low-power, and
noise-tolerant active optical imaging, and its fixed dwell time will
facilitate parallelization through use of a detector array.
\end{abstract}


\section{Introduction}
\label{introduction}
A light detection and ranging (LIDAR) system~\cite{schwarz2010mapping}
builds a histogram of photon counts over time.
The time delay and amplitude of the 
photon-count histogram, relative to the transmitted pulse's 
temporal profile,
contain object depth and reflectivity information, respectively, about the illuminated scene.
LIDAR signal-acquisition time must be long enough to collect the $10^2$ to $10^3$  photons per pixel (ppp)
needed to generate the finely-binned histogram required for accurate scene 3D and reflectivity images.

In this paper, 
we expound upon an active optical imaging framework that
recovers accurate reflectivity and 3D images simultaneously
using on the order of 1 detected ppp averaged over the scene.
This framework was introduced in \cite{ShinKGS:14icip} and builds upon an approach initiated
in~\cite{kirmani2014first}.
Like the first-photon imaging (FPI) method of~\cite{kirmani2014first},
our computational imager avoids the formation of histograms and instead uses
probabilistic modeling at the level of individual detected photons.

Both methods combine physically accurate probabilistic modeling of single-photon detection
with exploitation of the spatial correlations present in real-world scenes
to achieve
accurate 3D and reflectivity imaging when very little backreflected light reaches the detector,
as will be the case with low optical-power active imagers~\cite{mccarthy2009long}. 
The method introduced here uses deterministic dwell times,
which is both more convenient for raster scanning and
amenable to parallelization through the use of a detector array.
This ease of applicability comes with somewhat improved performance over FPI when compared at equal
total acquisition times in raster-scanned operation.

\subsection{Prior Work}
\label{prior_work}
\subsubsection{Active imaging methods} 
\begin{figure}
 \begin{center}
  \begin{tabular}{@{}c@{}}
   \includegraphics[width=0.48\textwidth]{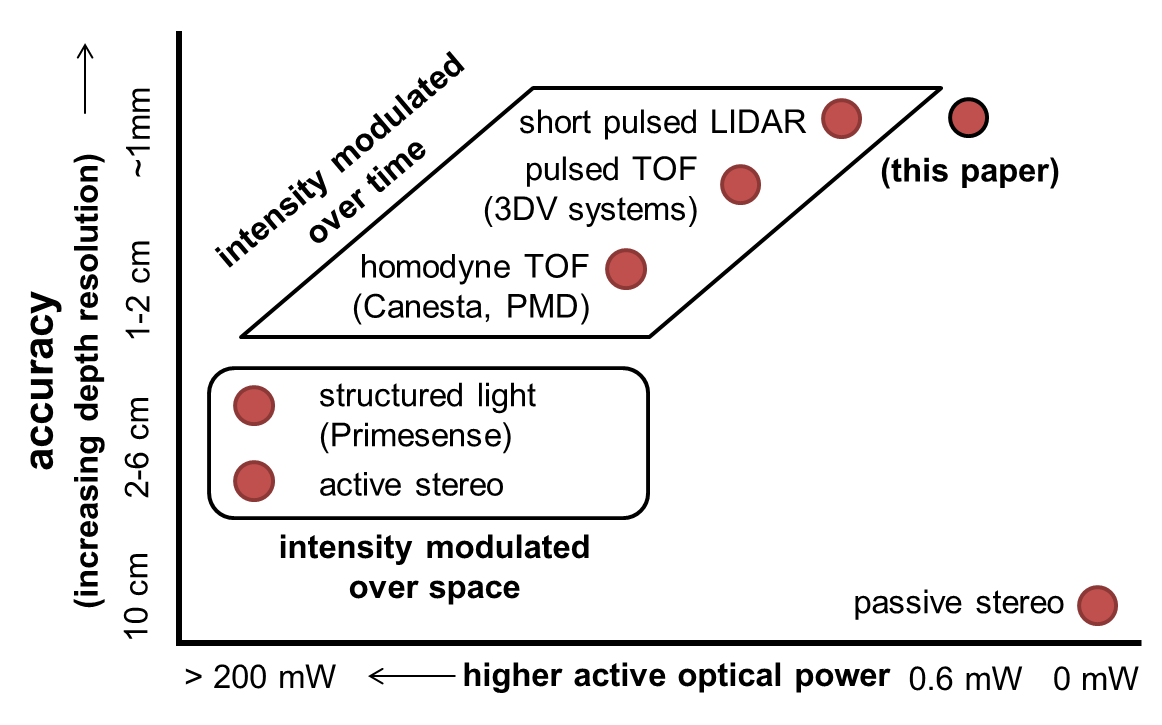} \\
   {\footnotesize (a) Accuracy vs.\ power trade-offs.} \\[2mm]
   \includegraphics[width=0.48\textwidth]{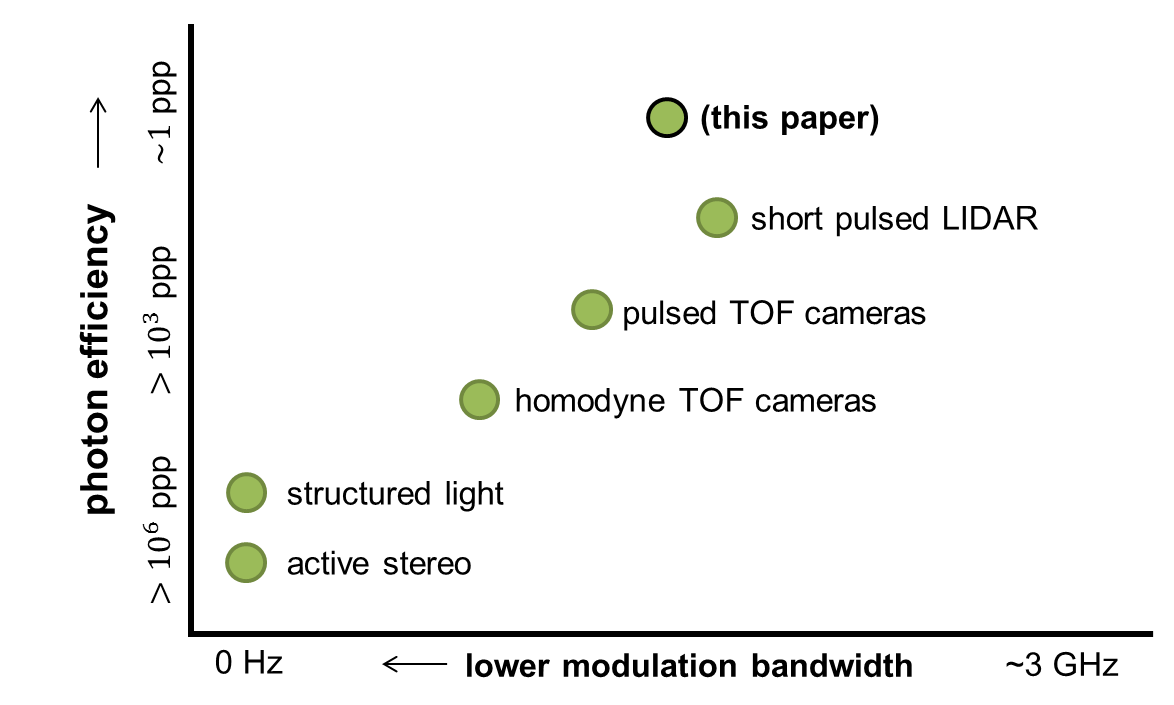} \\
   {\footnotesize (b) Photon efficiency vs.\ modulation bandwidth trade-offs.}
  \end{tabular}
 \end{center}
\caption{Qualitative comparison of state-of-the-art
  active optical 3D sensing technologies.
  Photon efficiency is defined as photons per pixel (ppp)
  necessary for centimeter-accurate depth imaging.}
\label{fig:prior}
\end{figure}

Active 3D imaging systems differ in how they modulate their transmitted power,
leading to a variety of trade-offs in accuracy, modulation frequency, optical power,
and photon efficiency; see Figure~\ref{fig:prior} for a qualitative summary.
Temporal modulation enables distance measurement by the time-of-flight (TOF) principle.
Examples of TOF acquisition systems,
ordered by increasing modulation bandwidth (decreasing pulse duration),
include homodyne TOF cameras~\cite{gokturk2004time},
pulsed TOF cameras~\cite{lee2013time}, and
picosecond laser radar systems~\cite{jelalian1980laser}. 
Spatial modulation techniques include
structured light~\cite{zhang2012microsoft} 
and active stereo imaging~\cite{forsyth2002computer}. 
These spatial-modulation techniques have low photon
efficiencies because they use an always-on optical source,
whereas pulsed-TOF systems have higher photon efficiencies
because they use sources that are on only for short intervals.
Additionally, the systems using temporal modulation have better accuracy than those
using spatial modulation.  The advantage of spatial modulation tends to be cheaper
sensing hardware, since high-speed sampling is not required.

The most photon-efficient TOF imagers---those requiring the fewest photons for
accurate imaging---use single-photon avalanche diode (SPAD) detectors~\cite{aull2002geiger}.
Earlier efforts in SPAD-based 3D imaging from
on the order of 1 detected ppp are
reported in~\cite{KirmaniVCWG:13-cleo,kirmani2013spatio,ShinKCG:13-GlobalSIP}.
The framework presented here
improves upon these works in part due to the use of estimated reflectivity.
This translates to SPAD-based imagers with lower optical power and
lower system bandwidth without sacrificing image quality.
There also has been significant recent interest in compressive methods
for 3D imaging, with~\cite{HowlandDH:11,ColacoKHHG:12-CVPR,HowlandLWH:13-OE}
and without~\cite{KirmaniCWG:11-OE} single-photon detection.
While compressive methods may reduce some measures of acquisition cost,
they do not generally improve photon efficiency.

\subsubsection{Optoelectronic techniques for low light levels}
In low light-level scenarios, 
a variety of optoelectronic techniques are employed for robust imaging.
Active imagers use lasers with narrow spectral bandwidths and spectral
filters to suppress background light
and minimize the Poisson noise it creates. 
However, optical filtering alone cannot completely eliminate background light,
and it also causes signal attenuation.
Range-gated imaging~\cite{busck2004gated} is another common technique,
but this method requires a priori knowledge of object
location. Furthermore, a SPAD may be replaced with a
superconducting nanowire single-photon detector (SNSPD)~\cite{goltsman2001picosecond},
which is much faster, has lower timing jitter, and has lower dark-count rate
than a SPAD\@. However, SNSPDs have much smaller active
areas and hence have narrower fields of view than SPAD-based systems with the same optics.

\subsubsection{Image denoising}
For depth imaging using SPAD data, it is typical to first
find a maximum likelihood (ML) estimate of scene depth
using a time-inhomogeneous Poisson process model for photon detection times
and then apply a denoising method.
The ML estimate is obtained independently at each pixel,
and the denoising is able to exploit the scene's spatial correlations.
This two-step approach commonly assumes a Gaussian noise model,
which is befitting because of the optimal behavior of ML with
large numbers of data samples.  However, at low light levels,
performing denoising well is more challenging due to the signal-dependent
nature of Poisson noise. In Section~\ref{experimental_results}, 
we compare our technique with the state-of-the-art denoising methods
that use sparsity-promoting regularization. Our superior
performance is due in part to our novel method for classifying
detection events as being due to signal (backscattered light)
or noise (background light and dark counts).

\subsubsection{First-photon imaging}
First-photon imaging~\cite{kirmani2014first}
is a method that allows accurate 3D and reflectivity reconstruction
using only the first detected photon at every pixel
in a raster-scanned scene.
FPI combines accurate first-photon detection statistics
with the spatial correlations existing in natural scenes
to achieve robust low light-level imaging.
The raster-scanning process of FPI, however, makes the dwell time
at each pixel a random variable.
Thus, FPI does not extend naturally to
operation using SPAD arrays---since simultaneous measurement implies equal dwell times---thus precluding 
the dramatic speedup in image acquisition that such arrays enable.

\begin{figure}
  \centering
  \centerline{\includegraphics[width=0.5\textwidth]{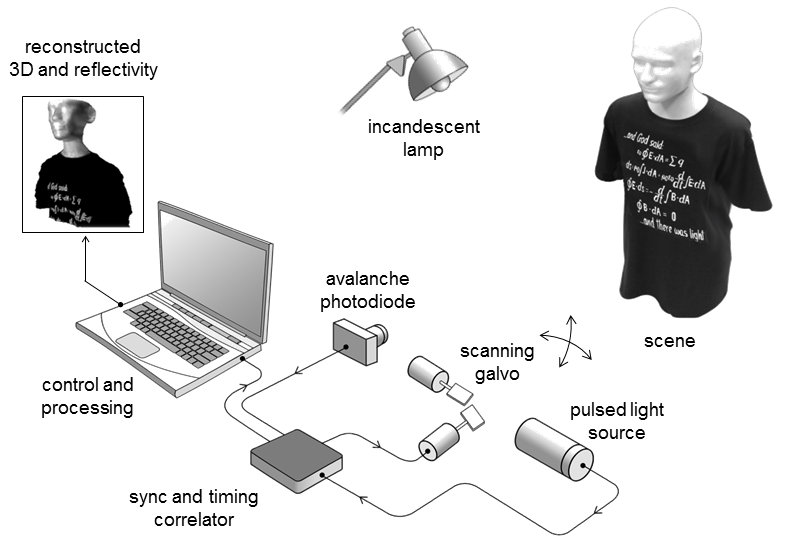}}
\caption{
Experimental imaging setup used with random dwell time in~\cite{kirmani2014first}
and with constant dwell time here.
A pulsed light source illuminates
the scene in a raster scan pattern. 
The backscattered 
light is collected by a time-resolved single-photon detector. 
Each spatial location is illuminated with exactly $N$ light pulses
(fixed dwell time).
An incandescent lamp injects background
light which corrupts the information-bearing signal. 
The photon detection times and
the total photon count are recorded at every image pixel. 
This dataset is used to estimate the 3D structure and reflectivity.
The setup is analogous to having a floodlight illumination source and an array of single-photon counting detectors
operating at a fixed dwell time.
}
\label{fig1}
\end{figure}

In this paper, 
we develop models and methods analogous to FPI that apply when there is a
fixed dwell time at each pixel.
In the experimental configuration depicted in Figure~\ref{fig1},
we demonstrate that the performance of the new method is similar to
or slightly better than FPI when compared for equal total acquisition time in raster-scanned operation.
Furthermore, with an $M$-fold increase in laser power and an $M$-element SPAD array, our fixed dwell-time framework can provide this same robust imaging $M$-times faster than a single-detector raster-scanned system.  

\subsection{Main contributions}

\subsubsection{Modeling}
We introduce a physically accurate model for the signal produced by a
SPAD under low light-level conditions that incorporates
an arbitrary illumination pulse shape,
background (ambient) light contribution,
dark counts, and the inhomogeneous
Poisson process characteristics (shot noise from the
quantum nature of light) given a fixed acquisition time.

\subsubsection{Algorithmic}
We provide a method for computational reconstruction of depth and reflectivity from
noisy photon-detection data. 
Our technique combines a shot-noise model for single photon-counting
statistics with the high degree of spatial correlation present in real-world scenes.

\subsubsection{Experimental}
We experimentally demonstrate 
that our proposed 3D imager's photon efficiency
is more than 100 times higher than that of traditional ML estimation.
We also show that our 3D imager
achieves sub-pulse-width depth resolution
under short acquisition times, in which
54\% of the pixels have missing data (no photon detections),
and
at high background levels, when
any given photon detection has approximately probability 0.5 of originating from ambient light.

\subsection{Outline}
The remainder of the paper is organized as follows.
Section~\ref{imaging_setup} introduces the LIDAR-like imaging configuration that we consider.
The key probabilistic models for the measured data are derived in Section~\ref{observation_model}.
These models are related to conventional image formation in Section~\ref{conventional},
and they are the basis for the novel image formation method in Section~\ref{novel}.
Section~\ref{experimental_results} presents experimental results for the novel method,
and
Section~\ref{conclusion} provides additional discussion and conclusions.
An appendix presents performance bounds based on our modeling.

The methods detailed in this paper were initially presented in abbreviated form in~\cite{ShinKGS:14icip}.
The present manuscript provides additional context, details on derivations, performance bounds,
and many more experimental results.

\section{Imaging Setup}
\label{imaging_setup}

\begin{figure*}
  \centering
  \centerline{\includegraphics[width=0.9\textwidth]{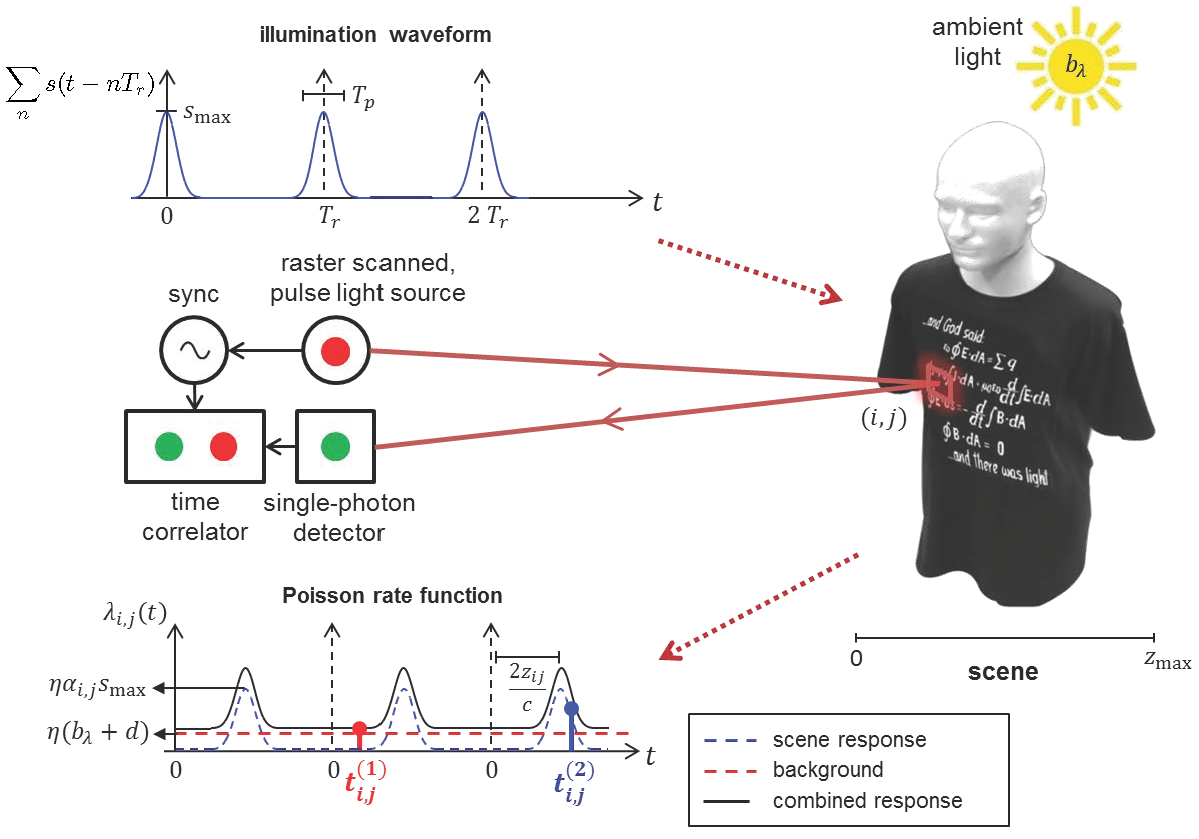}}
\caption{Summary of observation model.
Rate function of inhomogeneous Poisson process combining 
desired scene response and noise sources
is shown. 
Here, 
$N=2$ and $k_{i,j} = 2$.
A background count (red) occurred after the
second pulse was transmitted,
and a signal count (blue) occurred after
the third pulse was transmitted.
}
\label{fig:model}
\end{figure*}

Figure~\ref{fig:model} depicts the signal-acquisition model underlying our imager.
We aim to form reflectivity image $\balpha \in \mathbb{R}_+^{n \times n}$ and
depth image $\bz \in \mathbb{R}_+^{n \times n}$ of the scene.
We index the scene pixels as $(i,j)$, where $i,j = 1,\,2,\,\ldots,\,n$. 
The distance to pixel $(i,j)$ is $z_{i,j} \geq 0$
and its reflectivity, $\alpha_{i,j} \geq 0$, includes the effect of radial
fall-off, view angle, and material properties. 

\subsection{Illumination}
We use a periodically pulsed laser to illuminate the scene in a raster-scanned manner.
The repetition period is $T_r$ and the waveform of a single pulse is denoted by $s(t)$.
Physically, $s(t)$ is the photon-flux waveform of the
pulse emitted at $t=0$ measured in counts/sec (cps).
To avoid distance aliasing,
we assume $T_r > 2z_\text{max}/c$, where $z_\text{max}$ 
is the maximum scene depth and $c$ is the speed of light. 
With conventional processing,
the root mean-square (RMS) pulse width 
$T_p$ governs the achievable depth resolution
in the absence of background light~\cite{erkmen2009maximum}. 
As is typically done
in depth imaging, we assume that $T_p \ll 2 z_\text{max}/c.$

\subsection{Detection}
A SPAD detector provides time-resolved
single-photon detections~\cite{aull2002geiger}, called \emph{clicks}.
Its quantum efficiency $\eta$ is the
fraction of photons passing through the pre-detection optical filter that are detected.
Each detected photon is time stamped within a time
bin of duration $\Delta$, measuring a few picoseconds, that is much shorter than $T_p$.

\subsection{Data Acquisition}
Each pixel $(i,j)$ is illuminated with $N$ laser pulses.
The total acquisition time (dwell time) is thus $T_a = NT_r$.
We record the total number of photon detections $k_{i,j}$,
along with their detection times $\{t_{i,j}^{(\ell)}\}_{\ell=1}^{k_{i,j}}$, 
where the latter are measured relative to the immediately-preceding transmitted pulse. 
We also shine background light, with photon flux $b_\lambda$
at the operating optical wavelength $\lambda$, onto the detector.

\section{Probabilistic Measurement Model}
\label{observation_model}
Illuminating pixel $(i, j)$ with the
pulse $s(t)$ results in backreflected light with photon flux
$r_{i,j}(t) = \alpha_{i,j} s(t-2z_{i,j}/c)+b_\lambda$
at the detector.
The measurement of photon flux is through photon detections,
and carefully modeling the relationships between the measured quantities
and the reflectivity and depth variables is central to our imaging method.

\subsection{Poisson Statistics}
The photon detections produced by the SPAD in response to the backreflected light from transmission of $s(t)$ constitute an inhomogeneous Poisson process with time-varying rate function $\eta r_{i,j}(t)$.
To these photon detections we must add the detector dark counts, 
which come from an independent
homogeneous Poisson process with rate $d$.  Lumping the dark counts together with the background-generated counts yields the observation process at the SPAD's output, viz., 
as shown in Figure~\ref{fig:model}, an inhomogeneous Poisson process with rate function
\begin{eqnarray}
\lambda_{i,j}(t) 
&=& \eta  r_{i,j}(t)+d \nonumber \\
&=& \eta  \alpha_{i,j}  s(t-2z_{i,j}/c)+(\eta  b_\lambda+d),
\label{onepulserate}
\end{eqnarray}
when only a single pulse is transmitted.  Figure~\ref{fig:model} shows the rate function $\lambda_{i,j}(t)$ for the pulse-stream transmission.

Define $S = \int s(t) \,dt$ and $B = (\eta b_\lambda+d)T_r$ 
as the total signal and background count per pulse-repetition period, where we have used---and will use in all that follows---background counts to include dark counts as well as counts arising from ambient light.   
We assume that $B$ is known, because it is straightforward to
measure it before we begin data acquisition.
The derivations to follow assume $\eta \alpha_{i,j} S+B \ll 1$,
meaning that the photon-flux per pixel per pulse-repetition period is much less than 1,
as would be the case in low light-level imaging where an imager's photon efficiency is paramount.

\subsection{Distributions of Numbers of Detected Photons}
A SPAD detector is not \emph{number-resolving}, meaning that
it reports at most one click from detection of a signal pulse.
Using Poisson process properties~\cite{snyder1975random}, 
we have that the probability of the SPAD detector
\emph{not} recording a detection at pixel $(i,j)$
from one illumination trial is
\begin{align}
\label{zero}
P_0(\alpha_{i,j}) 
= \exp[ -(\eta  \alpha_{i,j}  S + B ) ].
\end{align}
Because we illuminate with a total of $N$ pulses,
and the low-flux condition ensures that multiple detections per repetition interval can be neglected,
the number of detected photons
$K_{i,j}$ is binomially distributed 
with probability mass function
\begin{eqnarray*}
\lefteqn{\text{Pr} \, [ K_{i,j}=k_{i,j};\alpha_{i,j} ] } \\
& = &
{N \choose k_{i,j}} \,
P_0(\alpha_{i,j})^{N-k_{i,j}} 
\left[
1-P_0(\alpha_{i,j}) 
\right]^{k_{i,j}},
\end{eqnarray*}
for $k_{i,j}=0,\,1,\,\ldots,\,N$.

In the ultimate low-flux limit in which $\eta \alpha_{i,j} S + B \rightarrow 0^+$
with $N \rightarrow \infty$
such that $N \{1-\exp [-(\eta \alpha_{i,j} S + B )]\} = C(\alpha_{i,j})$
is held constant,
$K_{i,j}$ converges to a Poisson random variable~\cite{bertsekas2002introduction}
with probability mass function 
\begin{align*}
\text{Pr}[  K_{i,j}=k_{i,j} ; \, \alpha_{i,j} ] 
=
\dfrac{
C(\alpha_{i,j})^k
}
{k \, !}
\,
\exp[-C(\alpha_{i,j})].
\end{align*}

\subsection{Distributions of Single-Photon Detection Times}
At pixel $(i,j)$, 
the single-photon detection time $T_{i,j}$ 
recorded by the SPAD detector 
is localized to a time bin of duration $\Delta$.
Because the SPAD detector only provides
timing information for the first (and, in the low-flux regime, \emph{only}) detected photon in a single pulse-repetition interval,
the probability of a SPAD click in $[t_{i,j},t_{i,j}+\Delta)$,
given there was a click in that repetition interval, is 
\begin{eqnarray*}
\lefteqn{
  \text{Pr} [\text{no click in $[0,t_{i,j})$, 
                click in $[t_{i,j},t_{i,j}+\Delta)$} \, 
                | \, \text{click in $[0,T_r)$} ]} \\
  & \!\!\overset{(a)}{=}\!\! &
\dfrac
{
\text{Pr}[ \text{no click in $[0,t_{i,j})$} ] \, \text{Pr}[ \text{click in $[t_{i,j},t_{i,j}+\Delta)$}] 
}
{
\text{Pr}[ \text{click in $[0,T_r)$}] 
}
\\
& \!\!\overset{(b)}{=}\!\! &
\dfrac{1}{1 - \exp[ -(\eta \alpha_{i,j} S + B)]}\,\,\times
\\
& &
\left\{
\exp \!\left[ -\int_0^{t_{i,j}} \! \left(\eta \alpha_{i,j} s\!\left(\tau-\frac{2z_{i,j}}{c}\right) + \frac{B}{T_r}\right) \, d\tau \right]
\right.
\\
& & \hspace{1mm} 
\left.
-\exp \!\left[ -\int_0^{t_{i,j}+\Delta} \! \left(\eta \alpha_{i,j} s\!\left(\tau-\frac{2z_{i,j}}{c}\right) + \frac{B}{T_r}\right) \, d\tau \right]
\right\}\!,
\end{eqnarray*}
where $(a)$ uses the independent increments property of the Poisson process and $(b)$ uses Equation~(\ref{zero}).
The probability density function of $T_{i,j} \in [0,T_r)$, the
continuous time-of-detection random variable, is then obtained by evaluating the preceding probability on a per unit time basis as $\Delta\rightarrow 0^+$:
\begin{eqnarray}
\lefteqn{f_{T_{i,j}}(t_{i,j};\, \alpha_{i,j},z_{i,j})}
\nonumber \\
& \!\!=\!\! &
\dfrac{1}{1-\exp[-(\eta  \alpha_{i,j}  S + B)]} \times \nonumber \\
& &
\lim_{\Delta\rightarrow 0^+}
\dfrac{1}{\Delta}
\left\{
\exp\!
\left[
-\int_0^{t_{i,j}} 
\left(\eta \alpha_{i,j} s\!\left(\tau-\frac{2z_{i,j}}{c}\right)
+\frac{B}{T_r} \right)
\, d\tau
\right]
\right.  \nonumber \\
& & \hspace{1mm}\left.
-
\exp\!
\left[
-
\int_0^{t_{i,j}+\Delta} 
\left(\eta \alpha_{i,j}
s\!\left(\tau-\frac{2z_{i,j}}{c} \right)
+\frac{B}{T_r}\right) 
\, d\tau
\right]
\right\}
\nonumber \\
& \!\!=\!\! &
\dfrac{
\eta \alpha_{i,j} s(t_{i,j}-2z_{i,j}/c)
+{B}/{T_r}}
{1-\exp[-(\eta \alpha_{i,j} S + B)]} 
\nonumber \\
& &
\,
\times \, \exp\!
\left[
- \int_0^{t_{i,j}}
\left(\eta \alpha_{i,j} s(\tau-2z_{i,j}/c)
+\frac{B}{T_r}\right) 
\, d\tau
\right]\nonumber \\
&\!\!\overset{(a)}{=}\!\! &
\dfrac{
\eta \alpha_{i,j} s(t_{i,j} -2z_{i,j}/c)
+B/T_r
}
{
\int_0^{T_r}
[\eta \alpha_{i,j} s(t_{i,j}-2z_{i,j}/c)
+B/T_r]
\, dt
} 
\nonumber \\
&\!\!\!=\!\!& 
\dfrac{\eta \alpha_{i,j} S}
{\eta \alpha_{i,j} S+B}
\left(  \dfrac{ s(t_{i,j}-{2z_{i,j}}/{c})}{S} \right)
+ 
\dfrac{B}
{\eta \alpha_{i,j} S+ B}
\left(
\dfrac{1}{T_r}
\right)\!\!,
\label{depth_pdf}
\end{eqnarray}
where $(a)$ follows from $\eta\alpha_{i,j}S + B \ll 1$.

A detection could be a signal count or a background count
The detection statistics result from the merging of the Poisson processes
corresponding to these sources. 
Under our low-flux assumption,
the detection time for a signal count from a single pulse-repetition interval is characterized by the normalized time-shifted 
pulse shape.
On the other hand, the detection time for a background count in that interval
is uniformly distributed on $[0,T_r)$.
Thus the probability density function in Equation~(\ref{depth_pdf})
is a mixture distribution, 
with mixture weights 
\begin{align*}
&
\text{Pr}[\,
\text{Detection at $(i,j)$ is signal}
\,] = 
\dfrac{\eta \alpha_{i,j} S}{\eta \alpha_{i,j} S + B}, 
\\
& 
\text{Pr}[\,
\text{Detection at $(i,j)$ is noise}
\,] = 
\dfrac{B}{\eta \alpha_{i,j} S + B}.
\end{align*}

\section{Conventional Image Formation}
\label{conventional}

\subsection{Pixelwise ML Reflectivity Estimation}
Given the total observed photon count $k_{i,j}$ at pixel $(i,j)$,
the constrained ML (CML) reflectivity estimate is
\begin{align*}
\hat \alpha_{i,j}^\text{CML} 
&= \argmax_{\alpha_{i,j}\geq 0 }
\text{Pr}[\, K_{i,j}=k_{i,j};\alpha_{i,j} \,] \\
&= 
\text{max}
\left\lbrace
\dfrac{1}{{\eta S}}
 \left[\log\!\left(\dfrac{N}{N-k_{i,j}}\right) -B \right],
\, 0
\right\rbrace.
\end{align*}
where $\log$ is the natural logarithm.
Traditionally, 
the normalized photon-count value
is used as the reflectivity estimate~\cite{chen1999photon},
\begin{align}
\label{eq_ml_ref}
\tilde \alpha_{i,j} = \dfrac{k_{i,j}}{N \eta S}.
\end{align}
Note that the normalized count value estimate 
is equal to the CML estimate under the Poisson approximation 
to the binomial distribution when $B=0$.

\subsection{Pixelwise ML Depth Estimation}
Using the 
photon detection-time dataset $\{t_{i,j}^{(\ell)}\}_{\ell=1}^{k_{i,j}}$,
the pixelwise constrained ML depth estimate is
\begin{align*}
\hat z_{i,j}^\text{CML} 
&= \argmax_{z_{i,j}\in [0,cT_r/2)} \prod_{\ell=1}^{k_{i,j}}
    f_{T_{i,j}}(t_{i,j}^{(\ell)}; \, \alpha_{i,j},z_{i,j}) \\
&= \argmax_{z_{i,j}\in [0,cT_r/2)} \sum_{\ell=1}^{k_{i,j}}
\log \! \left[
\eta \alpha_{i,j}  s\!\left(t_{i,j}^{(\ell)}-\frac{2z_{i,j}}{c}\right)
+ \frac{B}{T_r}
\right],
\end{align*}
assuming that $k_{i,j} \ge 1$.
If $B>0$, then the ML depth estimate is obtained
by solving a non-convex optimization problem.
Moreover, ML estimation when $B>0$ requires the knowledge
of the true reflectivity $\alpha_{i,j}$,
which is not typically available.
Thus, 
the log-matched filter~\cite{snyder1975random} is instead traditionally used 
for estimating depth from $k_{i,j}$ photon detections:
\begin{align}
\label{eq_ml_depth}
\tilde z_{i,j} =
\argmax_{z_{i,j}\in [0,cT_r/2)}
\sum_{\ell=1}^{k_{i,j}}
\log \!\left[
 s\!\left(t_{i,j}^{(\ell)}-{2z_{i,j}}/{c}\right)\right].
\end{align}
The log-matched filter solution is 
equal to the CML estimate when $B=0$.

\section{Novel Image Formation}
\label{novel}
In the limit of large sample size or high signal-to-noise ratio
(SNR), the ML estimate converges to the true parameter
value~\cite{kay1998fundamentals}. 
However, when the data is limited or SNR is low---such as in our problem---pixelwise ML solutions yield inaccurate
estimates. 
We compare our 3D imaging method with the baseline 
normalized-count reflectivity estimate $\tilde \alpha_{i,j}$
and the log-matched filter depth estimate $\tilde z_{i,j}$,
which are ML estimates asymptotically.
Along with using the single-photon detection statistics, 
we exploit the spatial correlations present
in real-world scenes
by regularizing the ML estimators.
Our approach provides 
significant improvements
over pixelwise ML estimators as well as 
traditional denoising techniques that may exploit scene sparsity
but assume additive Gaussian noise.
Our computational image formation proceeds in three steps.

\medskip
\noindent
\emph{Step 1: Reflectivity estimation.}
The negative log-likelihood of
scene reflectivity $\alpha_{i,j}$ given count data $k_{i,j}$ is
\begin{eqnarray} 
\lefteqn{\mathcal{L}_\alpha (\alpha_{i,j};k_{i,j}) = } \nonumber \\
&&
\hspace*{-.25in}(N-k_{i,j})  \eta  S  \alpha_{i,j}-
k_{i,j} \log\!\left\{ 1-
\exp [-(\eta  \alpha_{i,j}  S+B) ] \right\}\!, 
\end{eqnarray}
after constants independent of $\alpha_{i,j}$ are dropped.
Since $\mathcal{L}_\alpha (\alpha_{i,j};k_{i,j})$ is a 
strictly convex function in $\alpha_{i,j}$, it
is amenable to global minimization using convex optimization,
with or without the inclusion of 
sparsity-based regularization~\cite{harmany2012spiral}. 
The penalized ML (PML) estimate for scene reflectivity 
is obtained from noisy data $\{k_{i,j}\}_{i,j}$ by solving the
following convex program:
\begin{align*}
\hat \balpha^\text{PML} = 
\argmin_{\alpha:\alpha_{i,j} \ge 0}
\sum_{i=1}^n
\sum_{j=1}^n 
\,
\mathcal{L}_\alpha (\alpha_{i,j};k_{i,j})
+ \beta_\alpha \,\text{pen}_\alpha(\balpha),
\end{align*}
where $\text{pen}_\alpha(\cdot)$
is a convex function
that penalizes the non-smoothness of the reflectivity estimate,
and $\beta_\alpha$ controls the degree of penalization.

\medskip
\noindent
\emph{Step 2: Rejection of background detections.}
Direct application of a similar regularized-ML approach to
depth estimation using time-of-detection data is infeasible.
This is because the background contribution
to the likelihood function creates a nonconvex cost function
with locally-optimal solutions that are far from the global optimum.
Hence, before estimating depth,
a second processing step attempts to identify and censor the
detections that are due to background.

Background counts do not contain any scene-depth information.
Their detection times are mutually independent over spatial locations
with variance $T_r^2/12$.
In contrast, since light pulses have duration $T_p \ll T_r$
and depths $z_{i,j}$ are
correlated over spatial locations, the detection times of signal
counts have conditional variance, given data from
neighboring positions, that is much lower than $T_r^2/12$. Based on this key observation, our
method to censor a noisy detection at $(i, j)$ is as follows:
\begin{enumerate}[leftmargin=*]
\item Compute the rank-ordered mean (ROM)
$t^\text{ROM}_{i,j}$ for each pixel,
which is the median value of the detection times at the $8$
neighboring pixels of $(i,j)$~\cite{abreu1996new}.
If $t^\text{ROM}_{i,j}$ cannot be computed due to missing data, then set $t^\text{ROM}_{i,j} = \infty$.
\item Estimate the set of uncensored detections, $U_{i,j}$, i.e., those presumed to be signal detections, as follows:
\begin{align*}
\hspace*{-.1in}\left\lbrace
\ell :
| t_{i,j}^{(\ell)} - t^\text{ROM}_{i,j} | 
< 2 T_p  
\!\left(
\dfrac{ B }{ \eta  \hat\alpha^\text{PML}_{i,j} S + B }
\right)\!, 1 \leq\ell \leq k_{i,j}
\right\rbrace.
\end{align*}
\end{enumerate}
If $k_{i,j}=0$, then set $U_{i,j}=\varnothing$.
It is demonstrated in \cite{abreu1996new} that the method of rank-ordered means
is effective in detecting pixels that are corrupted by high-variance uniform noise.
Because background detections are uniformly distributed,
we use the ROM method to reject such detections and only keep signal detections for further processing.

\medskip
\noindent
\emph{Step 3: Depth Estimation.}
With background detections rejected, 
the negative log-likelihood function of depth $z_{i,j}$,
given uncensored data $\{t^{(\ell)}_{i,j}\}_{\ell \in U_{i,j}}$, 
is
\begin{align*}
\mathcal{L}_z \!\left(z_{i,j}; \, \{t_{i,j}^{(\ell)}\}_{\ell \in {U}_{i,j}} \right) 
= - \sum_{\ell \in U_{i,j}} \log\!\left[s\!\left(t_{i,j}^{(\ell)} - 2z_{i,j}/c\right)\right].
\end{align*}
If $|U_{i,j}|= 0$, then set
$\mathcal{L}_z (z_{i,j};\{t_{i,j}^{(\ell)}\}_{\ell \in {U}_{i,j}} )=0$,
so that it has no contribution to the scene's
negative log-likelihood cost function.

Our framework allows the use of arbitrary pulse shapes,
but many practical pulse shapes
are well approximated as $s(t) \propto \exp[-v(t)]$,
where $v(t)$ is a convex function in $t$. Then,
$\mathcal{L}_z  (z_{i,j};\{t_{i,j}^{(\ell)}\}_{\ell \in U_{i,j}} ) 
= \sum_{\ell \in U_{i,j}}  v (t_{i,j}^{(\ell)} - 2z_{i,j}/c )$
is a convex function in $z_{i,j}$.
Our penalized ML estimate
for the scene depth image is thus obtained using uncensored 
data and solving the following convex optimization problem:
\begin{align*}
\hat \bz^\text{PML} &=
\argmin_{z:z_{i,j} \in [0,cT_r/2)}
\sum_{i=1}^n
\sum_{j=1}^n
\mathcal{L}_z \!\left(z_{i,j};\{t_{i,j}^{(\ell)}\}_{\ell \in U_{i,j}} \right) \nonumber\\ 
&\hspace{.5in}+ \beta_z \, \text{pen}_z(\bz), \nonumber
\end{align*}
where ${\rm pen}_z(\cdot)$ is a convex function that penalizes
non-smoothness of the depth estimate, and $\beta_z > 0$
controls the degree of penalization.

\section{Experimental Results}
\label{experimental_results}
To test the performance of our proposed 
3D structure and reflectivity imaging method,
we used the dataset collected by D.~Venkatraman for~\cite{kirmani2014first},
which is available from~\cite{dheera}.
The experimental setup used to collect data is shown in Figure~\ref{fig1}. 
A pulsed laser diode 
with pulse width $T_p = 270 \, \text{ps}$ 
and repetition period $T_r = 100 \, \text{ns}$
was used as the illumination source.
A two-axis galvo was used to 
raster scan $1000 \times 1000$ pixels.
A lensless SPAD detector
with quantum efficiency $\eta = 0.35$ was used for detection.
The background light level was set such that $B$ equaled
the scene-averaged value of $\eta \alpha_{i,j} S$.
Further details of the experimental setup are given in~\cite{kirmani2014first}.
Because raster scanning with a fixed dwell time 
is equivalent to using a floodlight illumination source and a detector array,
our experimental results are indicative of what can be accomplished in
real-time imaging scenarios using SPAD arrays.

\subsection{Reflectivity Resolution Test}
\begin{figure*}
\centering
\begin{tabular}{c@{}c@{}c@{}c@{}c@{}c@{}c@{}c}
{}  
&
\hspace*{\newlengthtwo}
{\footnotesize (a) Photograph}
&
\hspace*{\mylength}
{\footnotesize (b) Ground truth} 
&
\hspace*{\mylength}
{\footnotesize (c) Pixelwise ML }
&
\hspace*{\mylength}
{\footnotesize (d) Denoising of (c) }
&
\hspace*{\mylength}
{\footnotesize (e) \text{Our method}}
&
\hspace*{\mylength}
{} 
&
\\
\rotatebox{90}{ \footnotesize \hspace{2mm} Reflectivity chart} 
&
\hspace*{\newlengthtwo}
\includegraphics[scale=0.50]{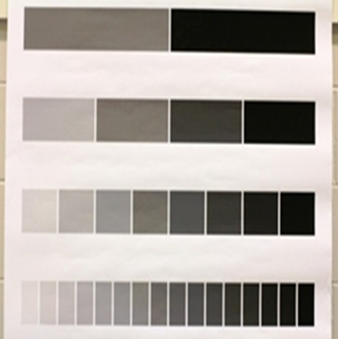} 
&
\hspace*{\mylength}
\includegraphics[scale=0.50]{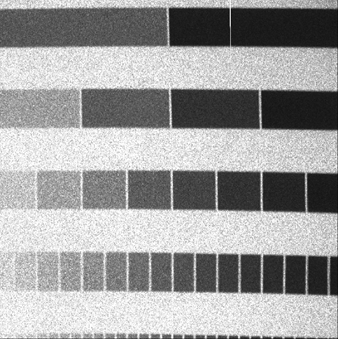} 
&
\hspace*{\mylength}
\includegraphics[scale=0.50]{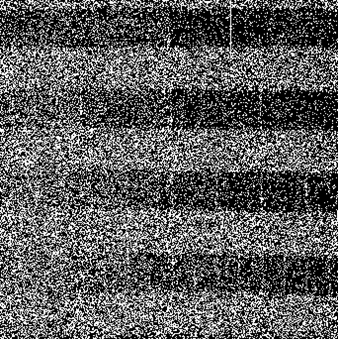} 
&
\hspace*{\mylength}
\includegraphics[scale=0.50]{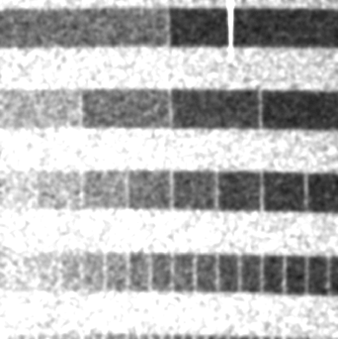} 
&
\hspace*{\mylength}
\includegraphics[scale=0.50]{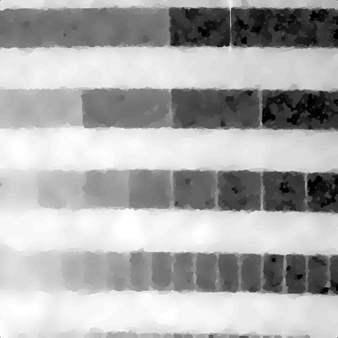} 
&
\hspace*{-3mm}
\hspace*{\mylength}
\includegraphics[scale=0.50]{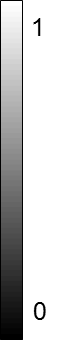} 
&
\vspace*{-1mm}
\\
{ } 
&
\hspace*{\newlengthtwo}
{\footnotesize } 
&
\hspace*{\mylength}
&
\hspace*{\mylength}
{\footnotesize $\text{PSNR} = 38.0\,\text{dB}$} 
&
\hspace*{\mylength}
{\footnotesize$\text{PSNR} = 51.3\,\text{dB}$} 
&
\hspace*{\mylength}
{\footnotesize$\text{PSNR} = 54.6\,\text{dB}$} 
&
\vspace{1mm}
\\
\rotatebox{90}{ \footnotesize \hspace{4mm} Depth chart} 
&
\hspace*{\newlengthtwo}
\includegraphics[scale=0.50]{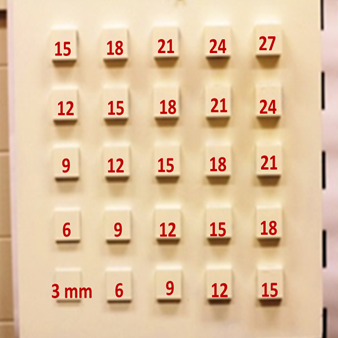} 
&
\hspace*{\mylength}
\includegraphics[scale=0.50]{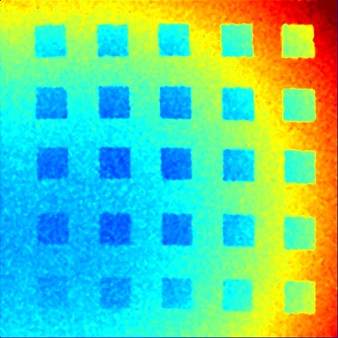} 
&
\hspace*{\mylength}
\includegraphics[scale=0.50]{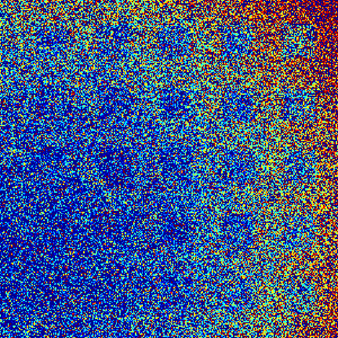} 
&
\hspace*{\mylength}
\includegraphics[scale=0.50]{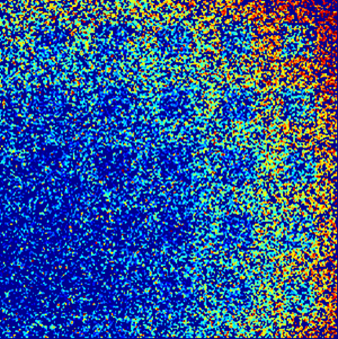} 
&
\hspace*{\mylength}
\includegraphics[scale=0.50]{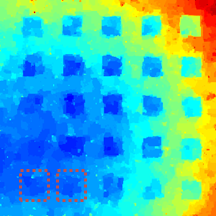} 
&
\hspace*{-0mm}
\hspace*{\mylength}
\includegraphics[scale=0.50]{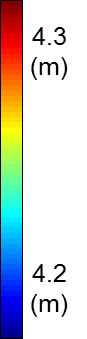} 
\vspace*{-1mm}
\\
{ } 
&
\hspace*{\newlengthtwo}
{\footnotesize } 
&
\hspace*{\mylength} 
&
\hspace*{\mylength}
{\footnotesize $\text{RMSE} = 322.8\,\text{cm}$} 
&
\hspace*{\mylength}
{\footnotesize$\text{RMSE} = 305.3\,\text{cm} $} 
&
\hspace*{\mylength}
{\footnotesize$\text{RMSE} = 0.4\,\text{cm} $} 
&
\hspace*{\barlength}
{\footnotesize } 
&
\end{tabular}
\caption{
Resolution test experiments.
Reflectivity chart imaging (top) was done using $T_a = 300 \,\mu\text{s}$
and had a mean count per pixel of 0.48\@.
They were scaled to fill the reflectivity interval $[0,1]$.
Depth chart imaging (bottom) was done using  $T_a = 6.2 \,\mu\text{s}$
and had a mean count per pixel of 1.1 with $33\%$
of the pixels having missing data, i.e., no detections.
}
\label{fig4}
\end{figure*}

Reflectivity estimation was
tested using the linear grayscale reflectivity chart shown in Figure~\ref{fig4}(a).  
Figure~\ref{fig4}(e) shows that our method resolves 
$16$ gray levels, performance similar
to that of the ground-truth image from Figure~\ref{fig4}(b), which required
about $1000$ photon detections per pixel. 

We quantified the performance of a reflectivity estimator $\hat \balpha$ of a true scene reflectivity $\balpha$
using peak signal-to-noise ratio (PSNR):
\begin{align*}
\text{PSNR}(\balpha,\hat\balpha)
= 10 \log_{10} 
\left( \dfrac{\text{max}_{i,j} \ \alpha_{i,j}^2 }{ \sum_{i,j} (\alpha_{i,j}-  \hat \alpha_{i,j})^2/n^2} \right)
\end{align*}
Figure~\ref{fig4} show that our method's PSNR exceeds that of pixelwise ML (Equation~(\ref{eq_ml_ref})) by $16$\,dB,
and it exceeds that of the bilateral-filtered~\cite{tomasi1998bilateral} pixelwise ML estimate by $3$\,dB.

\subsection{Depth Resolution Test}
Depth resolution was evaluated with
a test target comprising $5$\,cm $\times$ $5$\,cm squares of varying
thickness mounted on a flat board, as shown by the red-labeled squares in Figure~\ref{fig4}(a). The
smallest resolvable height (thickness) above the reference
level is an indicator of achievable depth resolution. 
Figure~\ref{fig4}(e) shows that our method achieves 4\,mm
depth resolution, which is comparable to that of the ground truth image (Figure~\ref{fig4}(b)), which required 100 detections per pixel, and far superior to the very noisy pixelwise ML image (Equation~(\ref{eq_ml_depth})), and its median-filtered~\cite{jain1995machine} version, which appear in Figures~\ref{fig4}(c) and (d), respectively.

We quantified the performance of a depth estimator $\hat \bz$
of a true scene depth $\bz$ using root mean-square error (RMSE):
\begin{align*}
\text{RMSE}({\bz},\hat {\bz})
= \sqrt{ \dfrac{1}{n^2} 
\sum_{i=1}^n \sum_{j=1}^n 
\, ( z_{i,j} - \hat z_{i,j} )^2}.
\end{align*}
At the background level in our experiment,
the pixelwise ML estimates 
have an RMSE of at least $3$\,m. 
Because many pixels are missing photon detection-time observations,
in order to denoise the pixelwise ML estimate,
we first perform bicubic interpolation
and then apply median filtering, which is effective in eliminating high-variance noise.
The depth resolution of
our method ($4$\,mm) corresponds to $760$-fold depth error
reduction, compared to the denoised estimate.

\subsection{Natural Scenes}
\begin{figure*}
\centering
\begin{tabular}{c@{}c@{}c@{}c@{}c@{}c@{}c@{}c@{}c}
{}  &
\hspace*{\newlengthtwo}
{\footnotesize (a) Ground truth}&
\hspace*{\mylength}
{\footnotesize (b) Pixelwise ML} &
\hspace*{\mylength}
{\footnotesize (c) Denoising of (b) }&
\hspace*{\mylength}
{\footnotesize (d) Our method }&
\hspace*{0.6mm}
{} &
\hspace*{\barlength}
{\footnotesize (e) RMSE images}&
\hspace*{1.7mm}
{} &
\\
\rotatebox{90}{ \footnotesize \hspace{4mm} Reflectivity} &
\hspace*{\newlengthtwo}
\includegraphics[scale=0.46]{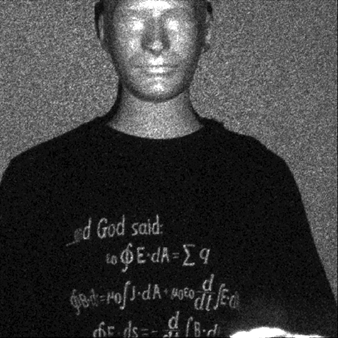} &
\hspace*{\mylength}
\includegraphics[scale=0.46]{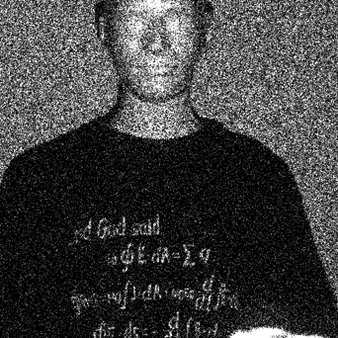} &
\hspace*{\mylength}
\includegraphics[scale=0.46]{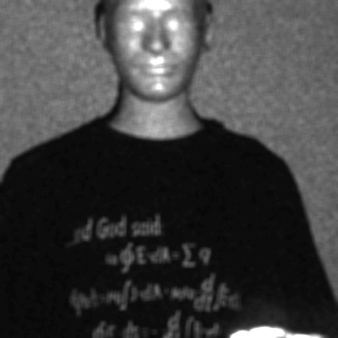} &
\hspace*{\mylength}
\includegraphics[scale=0.46]{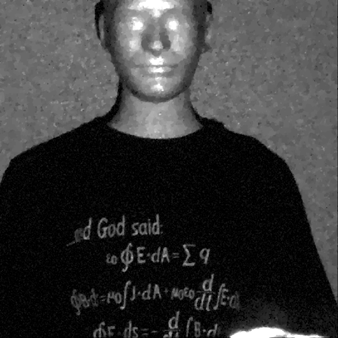} &
\hspace*{0.1mm}
\includegraphics[scale=0.46]{figs/i_bar} &
\hspace*{\barlength}
\includegraphics[scale=0.46]{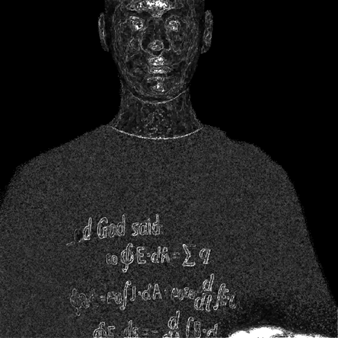} &
\hspace*{1.mm}
\includegraphics[scale=0.46]{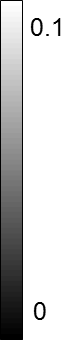} &
\vspace*{-1mm}
\\
{ } &
\hspace*{\newlengthtwo}
{\footnotesize } &
\hspace*{\mylength}
{\footnotesize $\text{PSNR} = 14.2\,\text{dB}$} &
\hspace*{\mylength}
{\footnotesize $\text{PSNR} = 26.5\,\text{dB}$} &
\hspace*{\mylength}
{\footnotesize$\text{PSNR} = 30.6\,\text{dB}$} &
\hspace*{\mylength}
\footnotesize  &
\hspace*{\barlength}
{\footnotesize } &
\vspace{0mm}
\\
\rotatebox{90}{ \footnotesize \hspace{4mm} 3D structure} &
\hspace*{\newlengthtwo}
\includegraphics[scale=0.46]{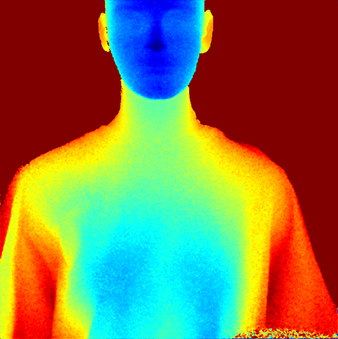} &
\hspace*{\mylength}
\includegraphics[scale=0.46]{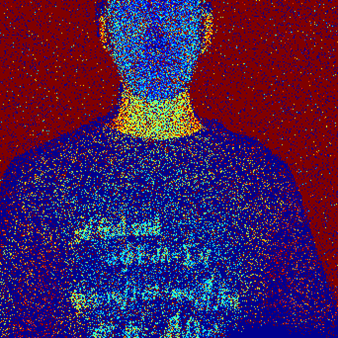} &
\hspace*{\mylength}
\includegraphics[scale=0.46]{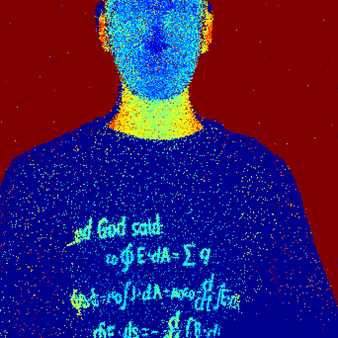} &
\hspace*{\mylength}
\includegraphics[scale=0.46]{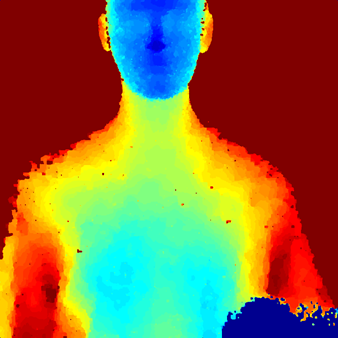} &
\hspace*{2.5mm}
\includegraphics[scale=0.46]{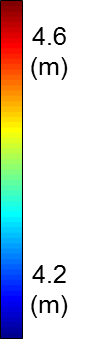} &
\hspace*{\barlength}
\includegraphics[scale=0.46]{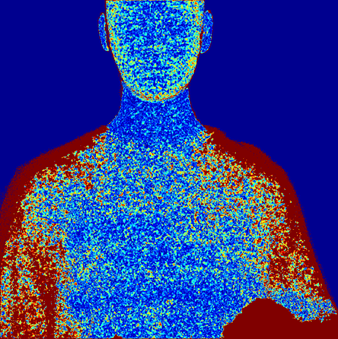} &
\hspace*{2mm}
\includegraphics[scale=0.46]{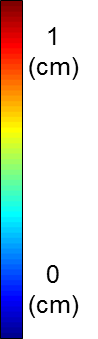} &
\vspace*{-1mm}
\\
{ } &
\hspace*{\newlengthtwo}
{\footnotesize } &
\hspace*{\mylength}
{\footnotesize $\text{RMSE} = 392.6\,\text{cm}$} &
\hspace*{\mylength}
{\footnotesize $\text{RMSE} = 362.8\,\text{cm}$} &
\hspace*{\mylength}
{\footnotesize$\text{RMSE} = 0.8 \,\text{cm} $} &
\hspace*{\mylength}
\footnotesize  &
\hspace*{\barlength}
{\footnotesize } &
\vspace{2mm}
\\
\rotatebox{90}{ \footnotesize \hspace{4mm} Reflectivity} &
\hspace*{\newlengthtwo}
\includegraphics[scale=0.46]{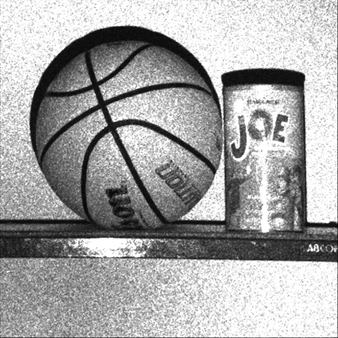} &
\hspace*{\mylength}
\includegraphics[scale=0.46]{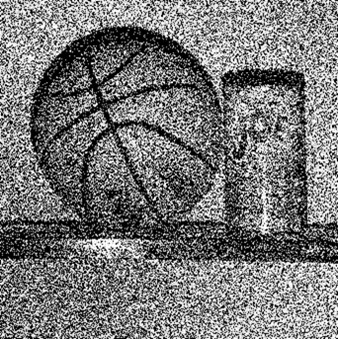} &
\hspace*{\mylength}
\includegraphics[scale=0.46]{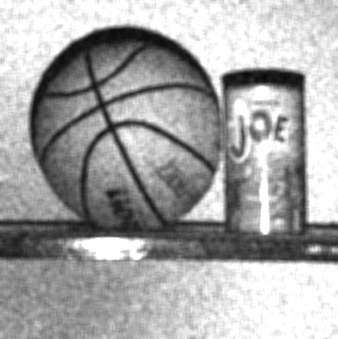} &
\hspace*{\mylength}
\includegraphics[scale=0.46]{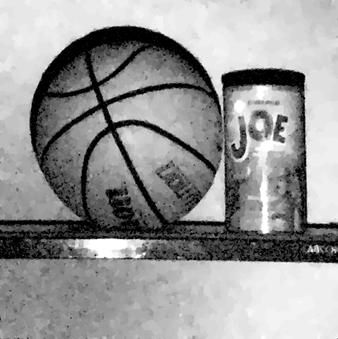} &
\hspace*{0.1mm}
\includegraphics[scale=0.46]{figs/i_bar} &
\hspace*{\barlength}
\includegraphics[scale=0.46]{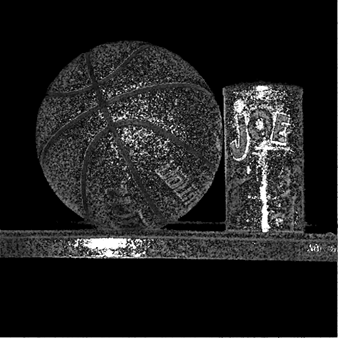} &
\hspace*{1.mm}
\includegraphics[scale=0.46]{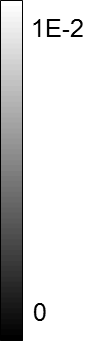} &
\vspace*{-1mm}
\\
{ } &
\hspace*{\newlengthtwo}
{\footnotesize } &
\hspace*{\mylength}
{\footnotesize $\text{PSNR} = 30.6\,\text{dB}$} &
\hspace*{\mylength}
{\footnotesize $\text{PSNR} = 42.3\,\text{dB}$} &
\hspace*{\mylength}
{\footnotesize$\text{PSNR} = 45.0\,\text{dB}$} &
\hspace*{\mylength}
\footnotesize  &
\hspace*{\barlength}
{\footnotesize } &
\vspace{0mm}
\\
\rotatebox{90}{ \footnotesize \hspace{4mm} 3D structure} &
\hspace*{\newlengthtwo}
\includegraphics[scale=0.46]{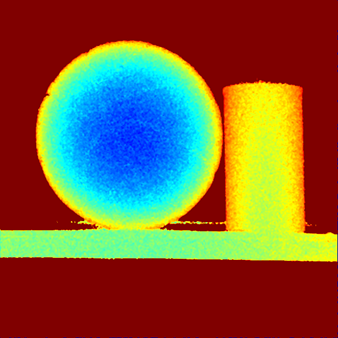} &
\hspace*{\mylength}
\includegraphics[scale=0.46]{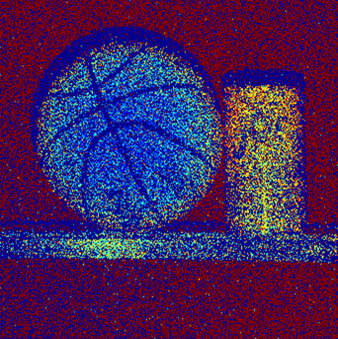} &
\hspace*{\mylength}
\includegraphics[scale=0.46]{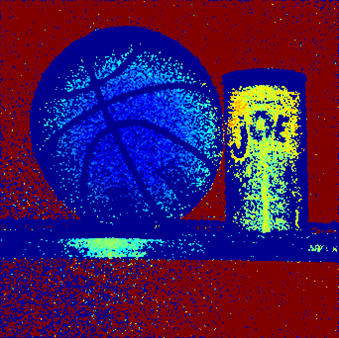} &
\hspace*{\mylength}
\includegraphics[scale=0.46]{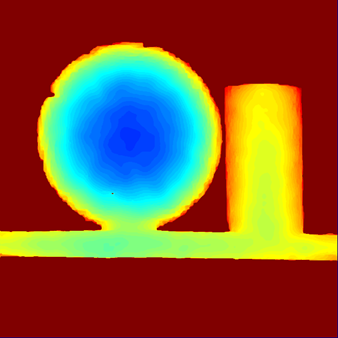} &
\hspace*{2mm}
\includegraphics[scale=0.46]{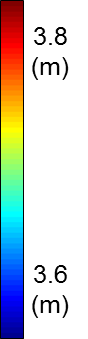} &
\hspace*{\barlength}
\includegraphics[scale=0.46]{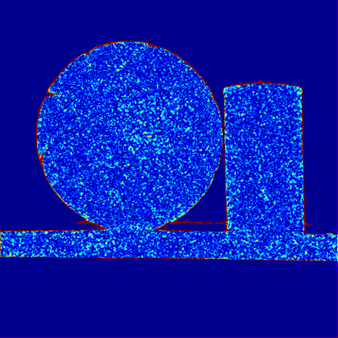} &
\hspace*{2.mm}
\includegraphics[scale=0.46]{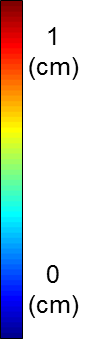} &
\vspace*{-1mm}
\\
{ } &
\hspace*{\newlengthtwo}
{\footnotesize } &
\hspace*{\mylength}
{\footnotesize $\text{RMSE} = 490.9\,\text{cm}$} &
\hspace*{\mylength}
{\footnotesize $\text{RMSE} = 460.8\,\text{cm}$} &
\hspace*{\mylength}
{\footnotesize$\text{RMSE} = 1.1\,\text{cm} $} &
\hspace*{\mylength}
\footnotesize  &
\hspace*{\barlength}
{\footnotesize } &
\end{tabular}
\caption{
Experimental results for reflectivity and 3D imaging of natural scenes.
We compare 
the reflectivity and depth images from
our proposed method
with those from pixelwise ML estimation (see Section~III).
Pixelwise RMSEs for the
reflectivity and 3D images 
using our method were generated
from $100$ trials of the experiments.
For the mannequin dataset (top), the mean per-pixel count was $1.2$ and $55\%$ of the pixels were missing data.
For the basketball-plus-can dataset (bottom), the mean per-pixel count was $2.1$ and $32\%$ of the pixels were missing data.
}
\label{fig5}
\end{figure*}

Reflectivity and depth images of two natural scenes---a life-size mannequin, and a basketball next to a can---are
shown in Figure~\ref{fig5}.  Ground-truth images, obtained using ML estimation from 200 detections at each pixel, appear in Figure~\ref{fig5}(a).  The mannequin dataset for pixelwise ML imaging and for our method was generated using acquisition time $T_a=100 \,\mu$s.  This dataset had $1.21$ detections per pixel averaged over the entire scene with 54\% of the pixels having no detections.  The basketball-plus-can dataset for pixelwise ML imaging and for our method also had $T_a = 100\,\mu$s, but its mean number of detections per pixel was 2.1, and 32\% of its pixels had no detections.  All reflectivity images were scaled to fill the interval $[0,1]$.

Figure~\ref{fig4}(b) shows that the pixelwise ML approach  
gives reflectivity and 3D estimates with 
low PSNR and high RMSE due to background-count shot noise at low light-levels.
Pixels with missing data
were imputed with the average of their 
neighboring $8$ pixelwise ML values.
Denoising the ML reflectivity estimate using bilateral filtering~\cite{tomasi1998bilateral}
and the ML depth estimate using median filtering~\cite{jain1995machine}
improves the image qualities (Figure~\ref{fig4}(c)).
However, denoising the 3D structure of the mannequin shirt fails,
because this region has very low reflectivity so that many of its pixels have missing data.
On the other hand, our framework, which combines accurate
photon-detection statistics with spatial prior information,
constructs reflectivity and 3D images with 30.6\,dB PSNR 
and $0.8$\,cm RMSE, respectively (Figure~\ref{fig4}(d)).
We used the total variation semi-norm~\cite{osher2003image}
as the penalty function in our method, and 
the penalty parameters were chosen to maximize PSNR for reflectivity imaging and minimize RMSE for 3D imaging.

\begin{figure}
\centering
\begin{tabular}{ c@{}c@{}c@{}c@{}c@{}c }
{} &
\hspace*{2mm}
\footnotesize  { LIDAR ($T_a=1000 \, \mu\text{s}$)}&
\hspace*{4mm}
\footnotesize  {Our method ($T_a = 30 \, \mu\text{s}$)}&
\\
\rotatebox{90}{ \footnotesize \hspace{4mm} Reflectivity} &
\hspace*{2mm}
\includegraphics[scale=0.55]{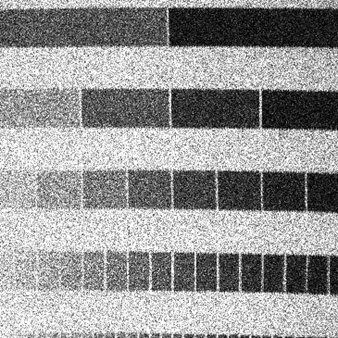} &
\hspace*{4mm}
\includegraphics[scale=0.55]{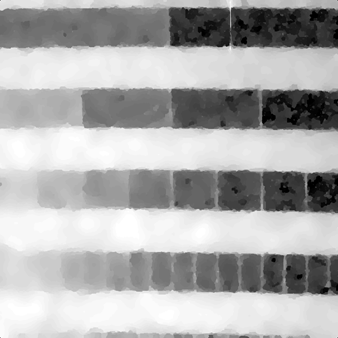} &
\\
{} &
\hspace*{2mm}
\footnotesize  {$\text{PSNR}= 52.7\,\text{dB}$}&
\hspace*{4mm}
\footnotesize  {$\text{PSNR}= 55.1\,\text{dB}$}&
\vspace{1mm}
\\
\rotatebox{90}{ \footnotesize \hspace{7mm} Depth} &
\hspace*{2mm}
\includegraphics[scale=0.55]{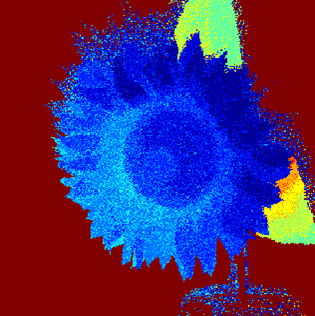} &
\hspace*{4mm}
\includegraphics[scale=0.55]{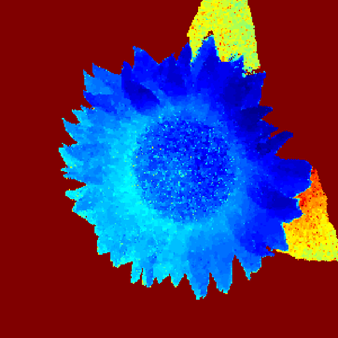} &
\\
{} &
\hspace*{2mm}
\footnotesize  {$\text{RMSE}= 0.59\,\text{cm}$}&
\hspace*{2mm}
\footnotesize  {$\text{RMSE}=0.41\,\text{cm}$}&
\end{tabular}
\caption{Comparison between our framework and conventional LIDAR technology.}
\label{fig8}
\end{figure}

Figure~\ref{fig8} shows how much photon efficiency
we gain over traditional LIDAR systems that use the histogramming approach.
The histogramming approach
is a pixelwise depth-estimation method
that simply searches for the location of the peak 
in the photon-count histogram of the backreflected pulse.
Whereas the log-matched filter is asymptotically ML as $B \rightarrow 0^+$,
histogramming-based depth estimation method
is asymptotically ML as $N\rightarrow \infty$.
Thus, when $T_a$ is long enough, as is the case in traditional LIDAR,
it is effective to use the histogramming-based depth estimation method.
Based on PSNR and RMSE values, 
we see that our framework can allow more than $30\times$
speed-up in acquisition, while constructing 
the same high-quality 3D and reflectivity images that a traditional LIDAR system 
would have formed using long acquisition times.

\subsection{Repeatability Test}
For each scene, we processed $100$ independent dataset
and computed the sample RMSE images 
that approximate
$\sqrt{\mathbb{E}[{(\alpha_{i,j}-\hat \alpha_{i,j}^\text{PML})^2}]}$
and 
$\sqrt{\mathbb{E}[{(z_{i,j}-\hat z_{i,j}^\text{PML})^2}]}$.
The pixelwise RMSE images, provided in Figure~\ref{fig5}(e),
corroborate the consistent accuracy and high resolution of 
our computational reflectivity and 3D imager.

\subsection{Effect of System Parameters}
\begin{figure}
\centering
\begin{tabular}{c@{}c@{}c}
\hspace*{-0.5cm}
\includegraphics[scale=0.3]{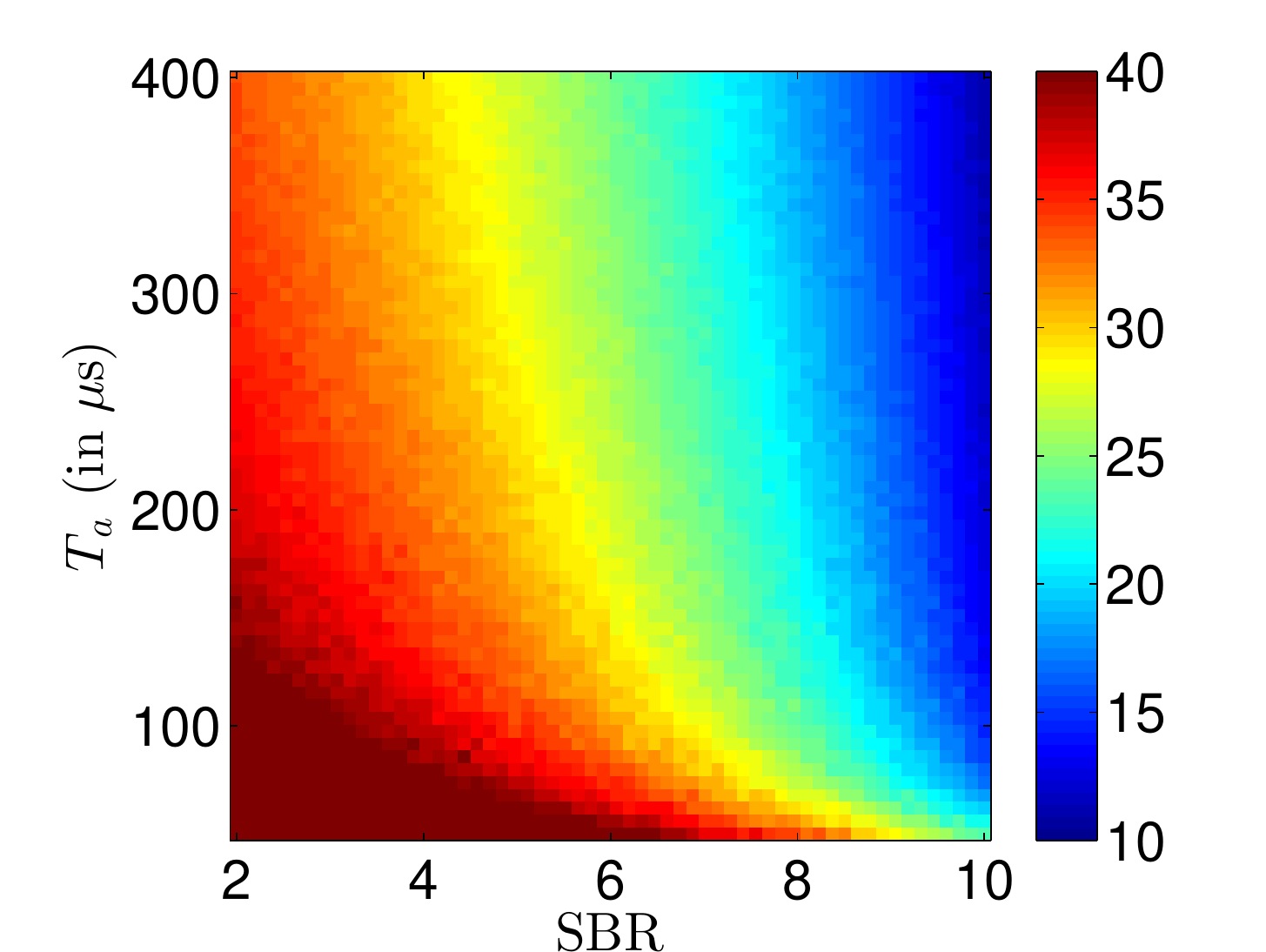} &
\includegraphics[scale=0.3]{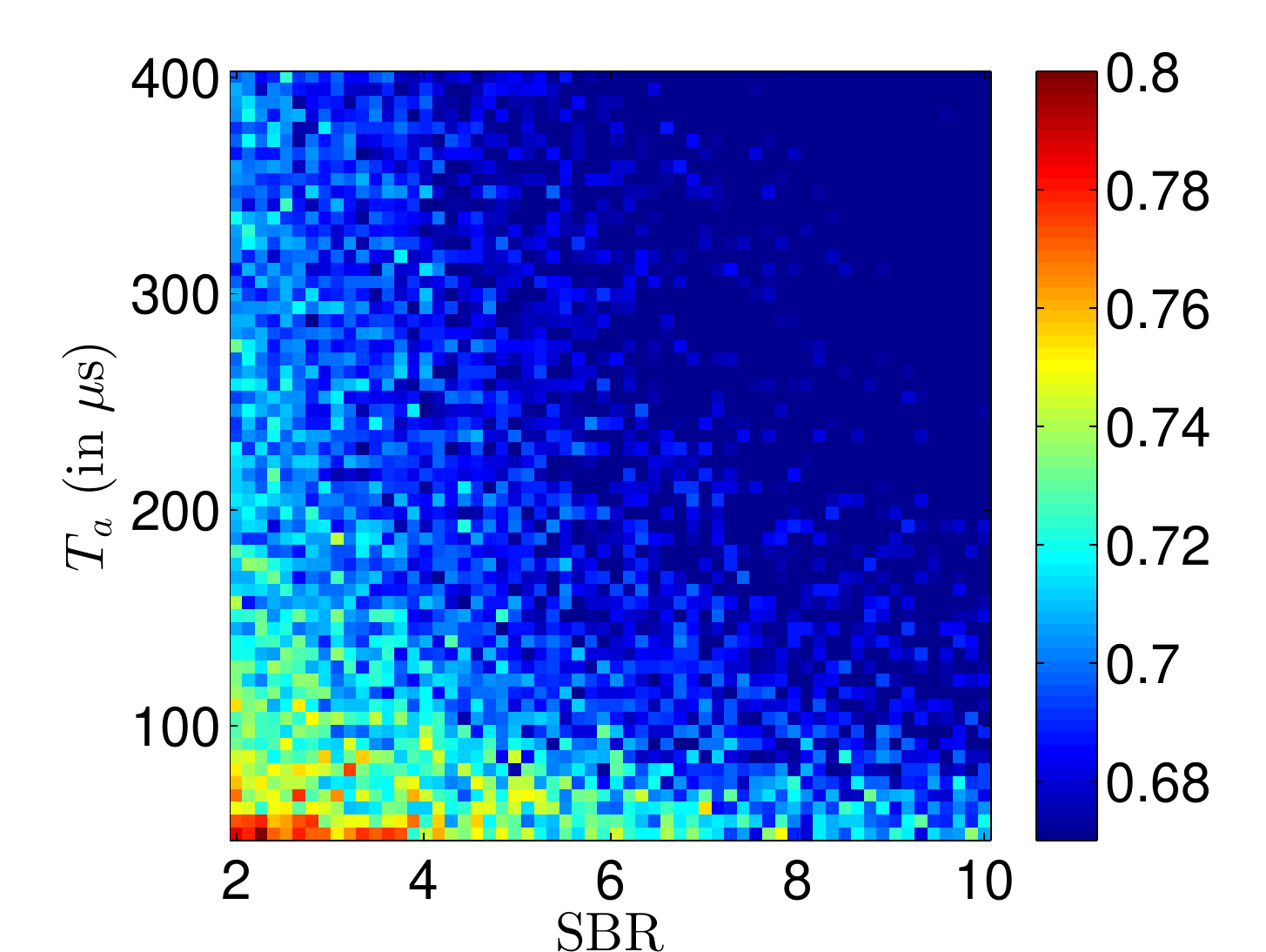} &
\\
\footnotesize {(a) RMSE (in cm) of}&
\footnotesize  {(b) RMSE (in cm) of}&
\vspace{-1mm}
\\
\footnotesize {pixelwise ML}&
\footnotesize  {our method}&
\end{tabular}
\caption{
RMSE results for 3D imaging.
SBR was varied by simulating background 
levels on the ground-truth mannequin dataset.
Note the differences in the colorbar scales.
}
\label{fig7}
\end{figure}

\begin{figure}
\centering
\begin{tabular}{@{}c@{\ \ }c@{\ \ }c@{\ \ }c@{}}
\rotatebox{90}{ \footnotesize \hspace{4mm} 3D structure} 
& \includegraphics[scale=0.6]{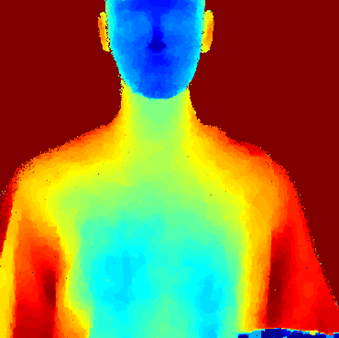} 
& \includegraphics[scale=0.6]{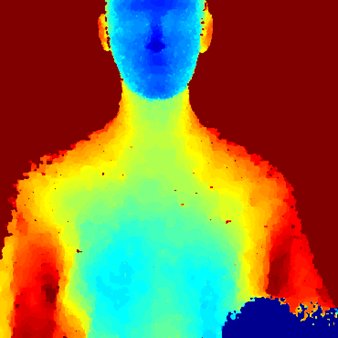} 
& \includegraphics[scale=0.6]{figs/man_d_bar} 
\\
& {\footnotesize (a) $T_a = 100\,\mu\text{s}$, $\text{SBR}=7$}
& {\footnotesize (b) $T_a = 100\,\mu\text{s}$, $\text{SBR}=1$}
\\[-1mm]
& {\footnotesize ($\text{RMSE} = 0.51\,\text{cm}$)} 
& {\footnotesize ($\text{RMSE} = 0.88\,\text{cm}$)} 
\\[3mm]
\rotatebox{90}{ \footnotesize \hspace{4mm} 3D structure} 
& \includegraphics[scale=0.6]{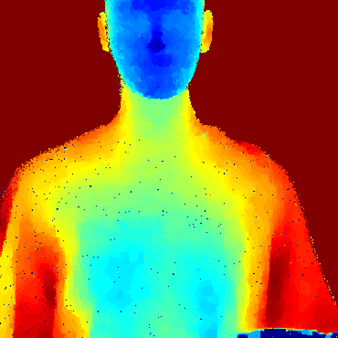} 
& \includegraphics[scale=0.6]{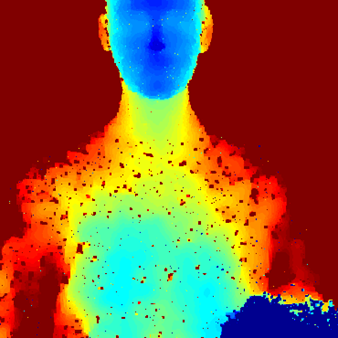} 
& \includegraphics[scale=0.6]{figs/man_d_bar} 
\\
& {\footnotesize (c) $T_a = 50\,\mu\text{s}$, $\text{SBR}=7$}
& {\footnotesize (d) $T_a = 50\,\mu\text{s}$, $\text{SBR}=1$}
\\[-1mm]
& {\footnotesize ($\text{RMSE} =  5.80\,\text{cm}$)} 
& {\footnotesize ($\text{RMSE} = 11.91\,\text{cm}$)} 
\\
\end{tabular}
\caption{
Effect of dwell time $T_a$ and signal-to-background ratio (SBR) 
on our 3D recovery method.
For acquisition times of $100\,\mu\text{s}$ and $50\,\mu\text{s}$, 
we calculated the mean photon count $k_{i,j}$ over all pixels to be $1.4$ and $0.6$, respectively. 
}
\label{fig6}
\end{figure}

Figure~\ref{fig7} shows how the performance of traditional ML and our image-formation methods
are affected by changing the
acquisition time $T_a$
and the signal-to-background ratio (SBR), 
defined to be
\begin{align*}
\text{SBR} = 
\dfrac{1}{n^2}
\sum_{i=1}^n
\sum_{j=1}^n
\dfrac{\eta \alpha_{i,j} S}{B}.
\end{align*}
In our experiment, SBR was modified by changing $T_r$
such that $B=(\eta b_\lambda +d) T_r$ is varied at constant $S$.
Figure~\ref{fig6} provides additional evidence that our method's
RMSE decreases monotonically with increasing $T_a$ and SBR, as one would expect.
More importantly, it shows that our 3D recovery method is robust under
strong background noise and short acquisition times.

\newlength{\mylengthh}
\setlength{\mylengthh}{5mm}

\subsection{Comparison with First-Photon Imaging}
First-photon imaging~\cite{kirmani2014first}
requires a single detection at each pixel, hence its dwell time on each pixel is a random variable.  Our method requires a fixed dwell time on each pixel, hence the number of detections on each pixel is a random variable.
So, to compare the performance of first-photon imaging with that of our method, we set the average per pixel dwell time of the former equal to the fixed per pixel dwell time of the latter.   That comparison, shown in Table~\ref{table2}, between the PSNRs of their reflectivity images and the RMSEs of their depth images, reveals several interesting characteristics.  In particular, when our method's image-acquisition time is matched to that of first-photon imaging, a substantial fraction of its pixels have missing data (no detections).  Nevertheless, our method successfully deals with this problem and yields performance similar to, or slightly better than, 
that of first-photon imaging for the five different scenes we have measured.

\subsection{Limitations}
Our method's depth image incurs its
highest error near the edges of scene objects. 
The surface normals
at these locations are nearly perpendicular to the line
of sight, which dramatically reduces SNR. Consequently,
these regions have fewer detections, with more of them being background counts, than do the rest
of the pixels. Although our method censors depth anomalies
near edges, it estimates the missing depths using spatial
correlations, leading to loss of subtle depth details.
Also, a detected photon may have originated from a multiple reflection, causing estimation inaccuracy. However, for quasi-Lambertian scenes, diffuse scattering causes multiple reflections to be considerably weaker than the direct
reflection. Combined with Poisson statistics, this implies an
exponentially diminishing probability of photon detection
originating from multiple reflections
Finally, our method of estimating reflectivity fails if the
background is sufficient to provide a detection in each
pulse repetition period with high probability. 
Hence, in our experiments, we employed a suitably narrowband spectral filter 
so that $\eta \alpha_{i,j} S \approx B \ll 1$ averaged over the scene.

\begin{table}
\centering
\begin{tabular}{l*{6}{l}r}
\hline
&  & \textbf{FPI} & \textbf{Ours}    \\
\hline
\hline
\multirow{5}{*}{Mannequin} 
& Mean $T_a$   & 244\,$\mu\text{s}$ & 244\,$\mu\text{s}$     \\
& Mean $k_{i,j}$   & 1\,ppp & 2.7\,ppp     \\
& Pixels missing data   & 0$\%$ & 33$\%$     \\
& PSNR  &  35\,dB &  37\,dB     \\
& RMSE    &   0.4\,cm &  0.3\,cm     \\
\hline
\multirow{5}{*}{Sunflower} 
& Mean $T_a$   & 15\,$\mu\text{s}$ & 15\,$\mu\text{s}$     \\
& Mean $k_{i,j}$  & 1\,ppp & 8.7\,ppp     \\
& Pixels missing data   & 0$\%$ & 18$\%$     \\
& PSNR    & 47\,dB & 47\,dB     \\
& RMSE  &   0.8\,cm &  0.5\,cm     \\
\hline
\multirow{5}{*}{Basketball and can} 
& Mean $T_a$   & 181\,$\mu\text{s}$ & 181\,$\mu\text{s}$     \\
& Mean $k_{i,j}$   & 1\,ppp & 1.7\,ppp     \\
& Pixels missing data   & 0$\%$ & 24$\%$     \\
& PSNR   &  44\,dB & 45\,dB     \\
& RMSE   &    1.1\,cm &  1.1\,cm     \\
\hline
\multirow{4}{*}{Reflectivity chart } 
& Mean $T_a$   & 120\,$\mu\text{s}$ & 120\,$\mu\text{s}$     \\
& Mean $k_{i,j}$   & 1\,ppp & 1.7\,ppp     \\
& Pixels missing data   & 0$\%$ & 27$\%$     \\
& PSNR    & 54\,dB &   56\,dB     \\
\hline
\multirow{4}{*}{Depth chart } 
& Mean $T_a$   & 6.2\,$\mu\text{s}$ & 6.2\,$\mu\text{s}$     \\
& Mean $k_{i,j}$   & 1\,ppp & 1.1\,ppp     \\
& Pixels missing data   & 0$\%$ & 35$\%$     \\
& RMSE    &   0.4\,cm &  0.4\,cm     \\
\hline \\
\end{tabular}
\caption{
Comparison between first-photon imaging and our framework.  Note that
$k_{i,j}$ is fixed and $T_a$ per pixel is a random variable for FPI, whereas $k_{i,j}$ is a random variable and $T_a$ per pixel is fixed for our framework.
}
\label{table2}
\end{table}

\section{Conclusions and Future Work}
\label{conclusion}
We have extended the FPI framework from~\cite{kirmani2014first}---which has a random per-pixel dwell time, because it records exactly one detection for each pixel in the scene---to one that has a fixed dwell time per pixel, but records a random number of detections for each pixel in the scene.  Both systems combine physically accurate single-photon detection statistics with exploitation of the spatial correlations found in natural scenes.  Our new method's fixed dwell time, however, makes it compatible with detector arrays.  Hence it is significant that we demonstrated its ability to produce accurate reflectivity and depth images
using on the order of 1 detected photon per pixel averaged over the scene, 
even with significant background light and a substantial fraction of the pixels having no detections. 
This highly photon-efficient performance motivates
the development of accurate and low-power SPAD array-based 
3D and reflectivity imagers.
Current commercial CMOS-based depth
imagers, for example Kinect and TOF cameras, have significantly
impacted research in 3D imaging. 
These sensors offer high depth resolution, but their use is limited 
due to poor spatial resolution and high power consumption.  Our approach offers a potential route to solving these problems.

More generally, our framework 
can be used in a variety of low light-level imaging applications
using photon-counting detectors, such as
spatially-resolved fluorescence lifetime imaging (FLIM)~\cite{o1984time}
and high-resolution LIDAR~\cite{aull2002geiger}.
Our method naturally extends to 
imaging at a variety of wavelengths, making it suitable
for practical implementations.
Furthermore, future advances in optoelectronic methods can improve the accuracy
of our 3D and reflectivity imager.
In particular, it can benefit from improved background
suppression techniques~\cite{mccarthy2009long} and
range-gating methods~\cite{busck2004gated}.

\appendix

This appendix provides performance analyses for pixelwise estimation.
The Cram\'er-Rao lower bound (CRLB)
sets the limit on the mean-square error (MSE)
of an unbiased estimator of a parameter.
Let $x$ be a scalar continuous parameter in the
probability density function $f_Y(y;x)$ of random variable $Y$.
The CRLB for an unbiased estimator, $\hat{x}$, of the parameter $x$ based on observation of $Y$ is the inverse of the Fisher information $J(x)$~\cite{kay1998fundamentals}:
\begin{align}
\mathbb{E}[( x - \hat x)^2] &\geq \text{CRLB}(x) = J^{-1}(x) \nonumber \\
&= 
\left\{ \mathbb{E} \left[ \dfrac{d^2}{d^2 x} 
\left(-\log f_Y(y;x)\right) \right] \right\}^{-1}.
\end{align}
An unbiased estimator $\hat x$ is {efficient} if 
$ \mathbb{E}[( x - \hat x)^2] = \text{CRLB}(x).$

\subsection{Mean-Square Error of Reflectivity Estimation}

With some algebra,
the CRLB for estimating the reflectivity, $\alpha_{i,j}$, at pixel $(i,j)$
can be shown to be
\begin{align}
\label{crlb_binomial}
\text{CRLB}\left(\alpha_{i,j}\right)
&= 
\left\{
\mathbb{E} \! \left[ \dfrac{d^2}{d^2 \alpha_{i,j}} 
\left(-\log \text{Pr}[K_{i,j}=k;\alpha_{i,j}] \right)\right]
\right\}^{-1} 
\nonumber \\ 
&=
\left\{
\mathbb{E} \!
\left[
\dfrac
{k\eta^2S^2 
\exp \left[\eta\alpha_{i,j}S+B \right]
}
{\left(
\exp \left[\eta\alpha_{i,j}S+B \right]
-1\right)^2}
\right]
\right\}^{-1}  
\nonumber \\
&=
\dfrac{
\exp
\left[
\eta\alpha_{i,j} S + B
\right]-1
}
{N\eta^2S^2} \nonumber \\
&\approx \dfrac{
\eta\alpha_{i,j} S + B
}
{N\eta^2S^2},
\end{align}
where the approximation makes use of the low-flux condition.
As could easily be expected, increasing the number of pulse repetitions, $N$, collects more photons and hence
decreases the CRLB.

Note, however, that we cannot directly use the CRLB result to lower bound 
the mean-square error of the 
unconstrained ML reflectivity estimate $\hat \alpha_{i,j}^\text{ML}$
given by
\begin{align*}
\hat \alpha_{i,j}^\text{ML} = \dfrac{1}{{\eta S}}
\left[\log\! 
\left(\dfrac{N}{N-k_{i,j}}\right)-B
\right].
\end{align*}
This is because the ML estimate is biased, 
($\mathbb{E}[\hat \alpha^\text{ML}_{i,j}] \neq \alpha_{i,j}$):
\begin{align*}
\mathbb{E}\left[\hat \alpha^\text{ML}_{i,j}\right] 
&= \mathbb{E} \!
\left[ \dfrac{1}{\eta S}\log\!\left( \dfrac{N}{N-k_{i,j}}\right) -\dfrac{B}{\eta S} \right] \\
& = \dfrac{1}{\eta S}\log N
- \dfrac{1}{\eta S}\mathbb{E} \left[\log \left(N-K_{i,j}\right) \right]
-\dfrac{B}{\eta S}\\
&> \dfrac{1}{\eta S}\log N
- \dfrac{1}{\eta S} \log \left(N-\mathbb{E}[K_{i,j}]\right) 
-\dfrac{B}{\eta S} \\
&= \alpha_{i,j},
\end{align*}
where the strict inequality comes from Jensen's inequality
and the fact that the logarithm function is strictly concave.

When $N \rightarrow \infty$ and $\eta\alpha_{i,j} S+B \rightarrow 0^+$ with 
$N [1 - \exp(\eta \alpha_{i,j} S + B)]$ equal to a constant $C(\alpha_{i,j})$, 
the ML reflectivity estimate is
\begin{align}
\hat \alpha_{i,j}^\text{ML} = 
 \dfrac{k}{N \eta S } - \dfrac{B}{\eta S}.
\end{align}
In this case, the CRLB equals 
the MSE of the ML reflectivity estimate,
\begin{align*}
\text{CRLB}(\alpha_{i,j})=
\mathbb{E}
\left[ \left( \alpha_{i,j} - \hat \alpha_{i,j}^\text{ML} \right)^2 \right]
=\dfrac{1}{N}
\left(
\dfrac{\alpha_{i,j}}{\eta S} + \dfrac{B}{\eta^2 S^2}
\right),
\end{align*}
We see that the CRLB expression from Poisson likelihood 
is equal to the first-order Taylor expansion of the CRLB expression
of the exact binomial likelihood given by Equation~(\ref{crlb_binomial}).

Knowing that the ML solution in the limiting Poisson distribution is unbiased and efficient,
we conclude that the maximum likelihood reflectivity estimate $\hat \alpha_{i,j}^\text{ML}$
is efficient asymptotically as $(\eta \alpha_{i,j} S + B)\rightarrow 0^+$ and $N\rightarrow \infty$.

\subsection{Mean-Square Error of 3D Estimation}
We again assume that
$\eta \alpha_{i,j} S + B \rightarrow 0^+$ and $N \rightarrow \infty$
such that $N [1 - \exp(\eta \alpha_{i,j} S + B)]$  is a constant $C(\alpha_{i,j})$.
The CRLB for estimating the depth $z_{i,j}$ is then found to be
\begin{align}
\text{CRLB}(z_{i,j})  
&= 
\left\{
\mathbb{E}
\left[ 
\dfrac{d^2}{d^2 z_{i,j}} 
\left(-\log f_{T_{i,j}}(  \{t_{i,j}^{(\ell)}\}_{\ell=1}^{k_{i,j}} ;z_{i,j}) \right)
\right]
\right\}^{-1}  
\nonumber \\
&= 
\left\{
\mathbb{E}
\left[ -
\sum_{\ell=1}^{k_{i,j}} 
\dfrac{d^2}{d^2 z_{i,j}} \log f_{T_{i,j}}(t_{i,j}^{(\ell)};z_{i,j}) \right]
\right\}^{-1}  
\nonumber \\
&=  
\dfrac{1}{C(\alpha_{i,j})}
\left( 
\int_{0}^{T_r} 
\dfrac
{ \dot{p}(t;z_{i,j})^2 }
{p(t;z_{i,j})} \, dt
\right)^{-1},
\end{align}
where 
\begin{align}
p(t;z_{i,j}) = \dfrac{\lambda_{i,j}(t)}{\int_0^{T_r} \lambda_{i,j}(\tau)\,d\tau} \nonumber
\end{align}
with $\lambda_{i,j}(t)$ being the single-pulse rate from Equation~(\ref{onepulserate}), and $\dot{p}(t;z_{i,j})$ is its derivative with respect to time.

We can exactly compute the MSE expression for 
certain pulse waveforms.
For example,
if the illumination waveform is a Gaussian pulse
$s(t) \propto
\exp
\left[
t^2/2 T_p^2
\right]$, 
then using the unconstrained log-matched filter expression,
we get
\begin{align*}
\hat z^\text{ML}_{i,j}
= \argmax_{z_{i,j}}
\sum_{\ell=1}^{k_{i,j}} 
\, \log\!\left[s(t^{(\ell)}_{i,j}-2z_{i,j}/c)\right]
= \dfrac{c}{2}\!
\left(
\dfrac{\sum_{\ell=1}^{k_{i,j}} t_{i,j}^{(\ell)}}{k_{i,j}}
\right),
\end{align*}
given $k_{i,j} \geq 1$.
If $k_{i,j}=0$,
then a standard pixelwise data imputation is done
by making a uniformly random guess over the interval $[0,cT_r/2)$.
Assuming $B=0$,
the MSE expression can be written as 
\begin{eqnarray}
\label{mse_depth}
\lefteqn{ \mathbb{E}[(z_{i,j} - \hat z_{i,j}^\text{ML})^2] 
\ = \
\mathbb{E}_{K_{i,j}}\{\mathbb{E}[(z_{i,j} - \hat z_{i,j}^\text{ML})^2 \,|\, K_{i,j} ]\}}
\nonumber \\
&\!\!=\!\!& 
\sum_{k=0}^\infty  \
\dfrac{C^k(\alpha_{i,j})e^{-C(\alpha_{i,j})}  }{ k!} 
\mathbb{E}[(z_{i,j} - \hat z_{i,j}^\text{ML})^2 \,|\, K_{i,j} =k]
\nonumber \\
&\!\!=\!\!& 
e^{-C(\alpha_{i,j})}
\left[
\left(\dfrac{cT_r}{2}\right)^2 + \left( z_{i,j} - \dfrac{cT_r}{4}\right)^2  
\right. \nonumber \\
& & \qquad \qquad \quad
\left. + \sum_{k=1}^\infty  \
\dfrac{C^k(\alpha_{i,j}) }{ k!} 
\dfrac{1}{k}\left(\dfrac{cT_p}{2}\right)^2
\right]
\nonumber \\
&\!\!=\!\!& 
e^{-C(\alpha_{i,j})}
\left(
\underbrace{
\left(\dfrac{cT_r}{2}\right)^2 
+ \left( z_{i,j} - \dfrac{cT_r}{4} \right)^2  
}_\text{random guess error}
\right. \nonumber \\ & & \qquad \qquad \quad \left.
\hspace*{-.1in}+ 
\underbrace{
\left(\dfrac{cT_p}{2}\right)^2 \int_{0}^{C(\alpha_{i,j})}
 \dfrac{\exp[\tau] - 1}{\tau} \, d\tau
}_\text{pulse-width error}
\right)\!\!.
\end{eqnarray}
As $C(\alpha_{i,j}) \rightarrow \infty$, the 
pulse-width error term in MSE dominates
and $\hat z^\text{ML}_{i,j}$ becomes an efficient estimator.

\bibliographystyle{IEEEtran}
\bibliography{references}

\begin{thebibliography}{10}
\providecommand{\url}[1]{#1}
\csname url@samestyle\endcsname
\providecommand{\newblock}{\relax}
\providecommand{\bibinfo}[2]{#2}
\providecommand{\BIBentrySTDinterwordspacing}{\spaceskip=0pt\relax}
\providecommand{\BIBentryALTinterwordstretchfactor}{4}
\providecommand{\BIBentryALTinterwordspacing}{\spaceskip=\fontdimen2\font plus
\BIBentryALTinterwordstretchfactor\fontdimen3\font minus
  \fontdimen4\font\relax}
\providecommand{\BIBforeignlanguage}[2]{{%
\expandafter\ifx\csname l@#1\endcsname\relax
\typeout{** WARNING: IEEEtran.bst: No hyphenation pattern has been}%
\typeout{** loaded for the language `#1'. Using the pattern for}%
\typeout{** the default language instead.}%
\else
\language=\csname l@#1\endcsname
\fi
#2}}
\providecommand{\BIBdecl}{\relax}
\BIBdecl

\bibitem{schwarz2010mapping}
B.~Schwarz, ``{LIDAR}: Mapping the world in 3d,'' \emph{Nat. Phot.}, 2010.

\bibitem{ShinKGS:14icip}
D.~Shin, A.~Kirmani, V.~K. Goyal, and J.~H. Shapiro, ``Computational 3d and
  reflectivity imaging with high photon efficiency,'' in \emph{Proc. IEEE Int.
  Conf. Image Process.}, Paris, France, Oct. 1995, to appear.

\bibitem{kirmani2014first}
A.~Kirmani, D.~Venkatraman, D.~Shin, A.~Cola{\c{c}}o, F.~N.~C. Wong, J.~H.
  Shapiro, and V.~K. Goyal, ``First-photon imaging,'' \emph{Science}, vol. 343,
  no. 6166, pp. 58--61, 2014.

\bibitem{mccarthy2009long}
A.~McCarthy, R.~J. Collins, N.~J. Krichel, V.~Fern{\'a}ndez, A.~M. Wallace, and
  G.~S. Buller, ``Long-range time-of-flight scanning sensor based on high-speed
  time-correlated single-photon counting,'' \emph{Appl. Opt.}, vol.~48, no.~32,
  pp. 6241--6251, 2009.

\bibitem{gokturk2004time}
S.~B. Gokturk, H.~Yalcin, and C.~Bamji, ``A time-of-flight depth sensor ---
  system description, issues and solutions,'' in \emph{Proc. IEEE Conf. Comput.
  Vis. Pattern Recog.}, 2004.

\bibitem{lee2013time}
S.~Lee, O.~Choi, and R.~Horaud, \emph{Time-of-Flight Cameras: Principles,
  Methods and Applications}.\hskip 1em plus 0.5em minus 0.4em\relax Springer,
  2013.

\bibitem{jelalian1980laser}
A.~V. Jelalian, ``Laser radar systems,'' in \emph{EASCON'80; Electronics and
  Aerospace Systems Conference}, vol.~1, 1980, pp. 546--554.

\bibitem{zhang2012microsoft}
Z.~Zhang, ``Microsoft {Kinect} sensor and its effect,'' \emph{IEEE Multimedia},
  vol.~19, no.~2, pp. 4--10, 2012.

\bibitem{forsyth2002computer}
D.~A. Forsyth and J.~Ponce, \emph{Computer Vision: A Modern Approach}.\hskip
  1em plus 0.5em minus 0.4em\relax Prentice Hall, 2002.

\bibitem{aull2002geiger}
B.~F. Aull, A.~H. Loomis, D.~J. Young, R.~M. Heinrichs, B.~J. Felton, P.~J.
  Daniels, and D.~J. Landers, ``Geiger-mode avalanche photodiodes for
  three-dimensional imaging,'' \emph{Lincoln Lab. J.}, vol.~13, no.~2, pp.
  335--349, 2002.

\bibitem{KirmaniVCWG:13-cleo}
A.~Kirmani, D.~Venkatraman, A.~Cola{\c c}o, F.~N.~C. Wong, and V.~K. Goyal,
  ``High photon efficiency computational range imaging using spatio-temporal
  statistical regularization,'' in \emph{Proc. CLEO}, San Jose, CA, Jun. 2013,
  paper QF1B.2.

\bibitem{kirmani2013spatio}
A.~Kirmani, A.~Cola{\c{c}}o, D.~Shin, and V.~K. Goyal, ``Spatio-temporal
  regularization for range imaging with high photon efficiency,'' in \emph{SPIE
  Wavelets and Sparsity XV}, San Diego, CA, Aug. 2013, pp. 88\,581F--88\,581F.

\bibitem{ShinKCG:13-GlobalSIP}
D.~Shin, A.~Kirmani, A.~Cola{\c c}o, and V.~K. Goyal, ``Parametric {P}oisson
  process imaging,'' in \emph{Proc. IEEE Global Conf. Signal Inform. Process.},
  Austin, TX, Dec. 2013, pp. 1053--1056.

\bibitem{HowlandDH:11}
G.~A. Howland, P.~B. Dixon, and J.~C. Howell, ``Photon-counting compressive
  sensing laser radar for 3d imaging,'' \emph{Appl. Optics}, vol.~50, no.~31,
  pp. 5917--5920, Nov. 2011.

\bibitem{ColacoKHHG:12-CVPR}
A.~Cola{\c c}o, A.~Kirmani, G.~A. Howland, J.~C. Howell, and V.~K. Goyal,
  ``Compressive depth map acquisition using a single photon-counting detector:
  Parametric signal processing meets sparsity,'' in \emph{Proc. IEEE Conf.
  Comput. Vis. Pattern Recog.}, Providence, RI, Jun. 2012, pp. 96--102.

\bibitem{HowlandLWH:13-OE}
G.~A. Howland, D.~J. Lum, M.~R. Ware, and J.~C. Howell, ``Photon counting
  compressive depth mapping,'' \emph{Opt. Expr.}, vol.~21, no.~20, pp.
  23\,822--23\,837, Oct. 2013.

\bibitem{KirmaniCWG:11-OE}
A.~Kirmani, A.~Cola{\c c}o, F.~N.~C. Wong, and V.~K. Goyal, ``Exploiting
  sparsity in time-of-flight range acquisition using a single time-resolved
  sensor,'' \emph{Opt. Expr.}, vol.~19, no.~22, pp. 21\,485--21\,507, Oct.
  2011.

\bibitem{busck2004gated}
J.~Busck and H.~Heiselberg, ``Gated viewing and high-accuracy three-dimensional
  laser radar,'' \emph{Appl. Opt.}, vol.~43, no.~24, pp. 4705--4710, 2004.

\bibitem{goltsman2001picosecond}
G.~Goltsman, O.~Okunev, G.~Chulkova, A.~Lipatov, A.~Semenov, K.~Smirnov,
  B.~Voronov, A.~Dzardanov, C.~Williams, and R.~Sobolewski, ``Picosecond
  superconducting single-photon optical detector,'' \emph{Appl. Phys. Lett.},
  vol.~79, no.~6, pp. 705--707, 2001.

\bibitem{erkmen2009maximum}
B.~I. Erkmen and B.~Moision, ``Maximum likelihood time-of-arrival estimation of
  optical pulses via photon-counting photodetectors,'' in \emph{Proc. IEEE Int.
  Symp. Inform. Theory}, 2009, pp. 1909--1913.

\bibitem{snyder1975random}
D.~L. Snyder, \emph{Random Point Processes}.\hskip 1em plus 0.5em minus
  0.4em\relax Wiley, New York, 1975.

\bibitem{bertsekas2002introduction}
D.~P. Bertsekas and J.~N. Tsitsiklis, \emph{Introduction to Probability}.\hskip
  1em plus 0.5em minus 0.4em\relax Athena Scientific, 2002.

\bibitem{chen1999photon}
Y.~Chen, J.~D. M{\"u}ller, P.~T. So, and E.~Gratton, ``The photon counting
  histogram in fluorescence fluctuation spectroscopy,'' \emph{Biophys. J.},
  vol.~77, no.~1, pp. 553--567, 1999.

\bibitem{kay1998fundamentals}
S.~M. Kay, \emph{Fundamentals of Statistical Signal Processing, Volume I:
  Estimation Theory}.\hskip 1em plus 0.5em minus 0.4em\relax Prentice Hall PTR,
  1998.

\bibitem{harmany2012spiral}
Z.~T. Harmany, R.~F. Marcia, and R.~M. Willett, ``This is {SPIRAL-TAP}: Sparse
  {P}oisson intensity reconstruction algorithms---theory and practice,''
  \emph{IEEE Trans. Image Process.}, vol.~21, no.~3, pp. 1084--1096, 2012.

\bibitem{abreu1996new}
E.~Abreu, M.~Lightstone, S.~K. Mitra, and K.~Arakawa, ``A new efficient
  approach for the removal of impulse noise from highly corrupted images,''
  \emph{IEEE Trans. Image Process.}, vol.~5, no.~6, pp. 1012--1025, 1996.

\bibitem{dheera}
``First-photon imaging project,''
  \url{http://www.rle.mit.edu/first-photon-imaging/}.

\bibitem{tomasi1998bilateral}
C.~Tomasi and R.~Manduchi, ``Bilateral filtering for gray and color images,''
  in \emph{Proc. 6th Int. Conf. Comput. Vis.}, 1998, pp. 839--846.

\bibitem{jain1995machine}
R.~Jain, R.~Kasturi, and B.~G. Schunck, \emph{Machine Vision}.\hskip 1em plus
  0.5em minus 0.4em\relax McGraw-Hill New York, 1995, vol.~5.

\bibitem{osher2003image}
S.~Osher, A.~Sol{\'e}, and L.~Vese, ``Image decomposition and restoration using
  total variation minimization and the {$H^{-1}$} norm,'' \emph{Multiscale
  Model. Simul.}, vol.~1, no.~3, pp. 349--370, 2003.

\bibitem{o1984time}
D.~O'Connor, \emph{Time-correlated single photon counting}.\hskip 1em plus
  0.5em minus 0.4em\relax Academic Press, 1984.

\end{thebibliography}

\end{document}